\documentclass[cpp,a4paper,fleqn%
]{w-art}
\usepackage{times,cite,w-thm}
\usepackage{amsmath}
\usepackage{bm}
\def\la{\mathrel{\mathchoice%
{\vcenter{\offinterlineskip\halign{\hfil$\displaystyle##$\hfil\cr<\cr\sim\cr}}}
{\vcenter{\offinterlineskip\halign{\hfil$\textstyle##$\hfil\cr<\cr\sim\cr}}}
{\vcenter{\offinterlineskip\halign{\hfil$\scriptstyle##$\hfil\cr<\cr\sim\cr}}}
{\vcenter{\offinterlineskip\halign{\hfil$\scriptscriptstyle##$\hfil\cr<\cr\sim\cr}}}}}
\newcommand{\be}{\begin{equation}}
\newcommand{\ee}{\end{equation}}
\newcommand{\der}[2]{\frac{d{#1}}{d{#2}}}

\newcommand{\pd}[2]{\frac{{\partial}{#1}}{{\partial}{#2}}}

\newcommand{\dvrg}[1]{\mathrm{div}{#1}}

\theoremstyle{plain}

\theoremstyle{definition}

\usepackage[]{graphicx}
\begin{document}
\DOIsuffix{theDOIsuffix}
\Volume{46} \Month{01}
\Year{2010}
\pagespan{1}{18} 
\Receiveddate{30 July 2010}
\Accepteddate{10 November 2010} \Dateposted{xx December 2010} 



\title[Wave instabilities of a collisionless plasma]{Wave instabilities
 of a collisionless plasma in fluid approximation}


\author[N.S. Dzhalilov]{Namig S. Dzhalilov\inst{1,2,3,}%
  \footnote{\ \ E-mail:~\textsf{NamigD@mail.ru}}}
\address[\inst{1}]{Astrophysikalisches Institut Potsdam (AIP), An der
Sternwarte 16, D-14482 Potsdam, Germany}
\address[\inst{2}]{Pushkov Institute of
Terrestrial Magnetism, Ionosphere and Radio
     Wave Propagation of the Russian  Academy of Sciences,
(IZMIRAN), Troitsk City, Moscow Region, 142190 Russia}
\address[\inst{3}]{Shamakhy
Astrophysical Observatory of the Azerbaijan Academy of Sciences
(ShAO), Baku Az-1000, Azerbaijan}
\author[V.D. Kuznetsov]{Vladimir D. Kuznetsov\inst{2,}
\footnote{E-mail:~\textsf{KVD@izmiran.ru}}}
\author[J. Staude]{J\"urgen Staude\inst{1,}
 \footnote{Corresponding author: E-mail:~\textsf{jstaude@aip.de},
            Phone \& Fax: +49\,3327\,43596}.} 
\begin{abstract}
Wave properties and instabilities in a magnetized, anisotropic,
collisionless, rarefied hot plasma in fluid approximation are
studied, using the 16-moments set of the transport equations
obtained from the Vlasov equations. These equations differ from the
CGL-MHD fluid model (single fluid equations by Chew, Goldberger, and
Low \cite{Chew,Baran}) by including two anisotropic heat flux
evolution equations, where the fluxes invalidate the double
polytropic CGL laws. We derived the general dispersion relation for
linear compressible wave modes. Besides the classic incompressible
fire hose modes there appear four types of compressible wave modes:
two fast and slow mirror modes -- strongly modified compared to the
CGL model -- and two thermal modes. In the presence of initial heat
fluxes along the magnetic field the wave properties become different
for the waves running forward and backward with respect to the
magnetic field.
 The well known discrepancies between the results of the CGL-MHD
fluid model and the kinetic theory are now removed: i) The mirror
slow mode instability criterion is now the same as that in the
kinetic theory. ii) Similarly, in kinetic studies there appear two
kinds of fire hose instabilities - incompressible and compressible
ones. These two instabilities can arise for the same plasma
parameters, and the instability of the new compressible oblique fire
hose modes can become dominant. The compressible fire hose
instability is the result of the resonance coupling of three
retrograde modes - two thermal modes and a fast mirror mode. The
results can be applied to the theory of solar and stellar coronal
and wind models.
\end{abstract}
\maketitle                   




\renewcommand{\leftmark}
{N.S. Dzhalilov et al.: Wave instabilities of a collisionless
plasma}

\section{Introduction}

Frequent particle collisions turn the plasma distribution function
into an isotropic one, and thus the thermal pressure is isotropic as
well. If collisions rarely occur the presence of a magnetic field
will maintain a ``non-mixed'' state of the energies of the
longitudinal and transverse motions of particles. Thus, the
transverse and longitudinal kinetic particle temperatures will
differ from each other, $T_\bot\ne T_\|$ . Typical examples of such
plasmas are the coronal and solar wind plasmas which are very
anisotropic and inhomogeneous, in cross-field direction in
particular \cite{Asc05}. For the observational motivation of our
modeling see the references in our previous paper \cite{Dzhal08}.
So, due to the anisotropy of the kinetic temperatures of the
particles (especially protons and heavy ions) the corresponding
partial pressures become anisotropic in this way. This makes the
total thermal pressure anisotropic too, $p_\bot\ne p_\|$.

In the corona the electron and ion gyroradii $r_{B}$ and gyrotimes
$\tau_{B}$ become smaller than any of the particle collisional mean
free paths and times and smaller than any of the typical scales of
variations of macroscopic thermodynamical quantities. Thus, the
condition of a strongly magnetized plasma is well satisfied. That
means, particles gyrating around the magnetic field lines are
localized across the field at a distance of the Larmor radius which
plays the role of a free path length of particles. Thus, the
dynamical motion of a collisionless plasma with characteristic
scales of $L\gg r_{B}$ and $\tau\gg \tau_{B} $ behaves across the
magnetic field as a fluid. However, under such circumstances a
traditional hydrodynamical description of the plasma is hardly
possible \cite{Marsch06}. The isotropic MHD equations are applicable
only if the plasma is collision-dominated and the distribution
functions are close to Maxwellian's. In the opposite limiting case,
when the plasma is collisionless at all, the local distribution
function strongly differs from the Maxwell function. To describe the
plasma in the fluid approximation in this case usually the single
fluid CGL-MHD equations by Chew et al. \cite{Chew,Baran} are
applied. Instead of the energy equation of the isotropic MHD these
equations include double-polytropic laws in the form $p_\bot/\rho B
= $ const and $p_\| B^2/\rho^3=$ const. Many studies of wave
instability problems are based on these equations. Similar to the
low-frequency kinetic considerations there appear two kinds of
instabilities: incompressible fire hose and compressible mirror
instabilities
\cite{veden58,chandra58,parker0,barnes66,Hasegawa69,Gary98}. In
comparison to the results based on the kinetic theory the CGL
equations provide the correct instability criterion for the classic
fire hose instability. However, there exists a discrepancy in the
criterion for the slow-mode mirror instability. Moreover, there
appear basic differences between the nonlinear stages of these
instabilities compared with the kinetic results. It has been shown
\cite{Hellinger00,Hellinger01} by hybrid kinetic simulations that a
new type of fire hose instability may arise for oblique propagation
due to the proton temperature anisotropy. Unlike the classical fire
hose instability this instability is compressible and has a maximum
growth rate at oblique propagation. The growth rate of the new
instability is comparable to (or in some parameters ranges even
larger than) the maximum of the standard fire hose instability
growth rate. Both fire hose instabilities may occur at the same time
for the same plasma parameters. A similar second type of fire hose
instability driven by the electron temperature anisotropy has also
been found \cite{Hollweg70,Pae99,Li00}. The CGL theory cannot give
an analogy of this second type of fire hose instability.

To remove such discrepancies in the CGL-MHD model, generalized
polytropic laws were used \cite{Abraham73,Hau93,Wang03} introducing
some artificial polytropic indices such as  $p_\bot/\rho
B^{\gamma_1-1} = $ const and $p_\| B^{\gamma_2-1}/\rho^{\gamma_2} =$
const. With a suitable choice of these free polytropic indices,
$\gamma_1$ and $\gamma_2$, it is in principle possible to remove
some of the discrepancies. However, the origin of these new indices
is not clear as they do not follow directly from the kinetic
equations when the fluid equations are derived. Later it has been
shown that in the experimental data and in the particle simulations
these two adiabatic invariants become invalid for realistic plasmas
\cite{Marsch87,Quest96}. The conservation of the two CGL adiabatic
invariants in an ideal collisionless plasma leads to a strong
pressure anisotropy $p_\perp < p_\|$ which is much larger than the
observed values \cite{Grooker77,Hill95}. However, due to non-ideal
effects such as heat flux these invariants are broken \cite{Hau96},
and this leads to properties which are quiet different from those
predicted by the CGL equations. Deriving the CGL-MHD equations the
third moments of the distribution function, hence the heat fluxes
have been ignored without any proof \cite{Baran}, which is the main
shortage of these equations.

In the present paper we study the linear wave instability problem on
the base of more correct equations -- the 16-moments transport
equations, which are derived from the Vlasov collisionless
magnetized plasma kinetic equations by the fast gyromotion ordering
technique \cite{Oraev68,Oraev85,Ramos03}. These equations include
additionally two dynamic evolution equations of the heat fluxes and
no polytropic laws are possible. This allows us to resolve the main
discrepancies of the CGL fluid theory. We consider the wave
peculiarities which can appear in the anisotropic compressible
plasma. We have already considered the incompressible wave
instability on the base of these equations \cite{Dzhal08}. In
Section 2 we formulate the basic equations, which are the integrated
moment equations of the kinetic Vlasov equations. In Section 3 the
linear compressible wave equation and the general dispersion
relation are derived. To compare the results with the CGL-MHD theory
the CGL dispersion equation is deduced from the new dispersion
equation as a special case in Section 4. The solutions and analysis
of the new dispersion equation are the topic of Section 5. In
Section 6 we show that similar to the the kinetic theory the
existence of two kinds of fire hose instabilities is possible in the
fluid approximation too. The mass density fluctuations due to
compressible wave modes and their instability is discussed in the
Section 7. A discussion and some conclusions are presented in
Section 8.

\section{Basic equations}

A plasma is described by the system of kinetic equations for the
distribution functions of the particles and the Maxwell equations
for the electromagnetic field. Due to the complexity of the kinetic
equations the large-scale behavior of the plasma is usually
described by deducing the equations for the integrated moments of
the distribution function, and these equations are referred to as
the hydrodynamical or the transport equations. The set of usual MHD
equations is one variant of such equations valid for the
collision-dominated isotropic plasma. For the description of a
collisionless anisotropic plasma the 16--moments set of equations
may be used which is more complete including the evolution of heat
fluxes. This set of equations has been used by many authors in
different theoretical approaches, especially for modeling the
ionospheric plasma \cite{Oraev68,Oraev85} and the solar wind
\cite{Demars79,Olsen99,Li99,Lie01}. A more correct and compact form
of these single-fluid transport equations for the anisotropic plasma
in the presence of gravity $g$ but without magnetic diffusivity
under the conditions $r_B\ll V \tau$ and $r_B\ll v_T \tau$ has been
derived \cite{Oraev85}; see \cite{Oraev68,Ramos03}. These equations
are given as follows
\begin{eqnarray}
&&\der{\rho}{t}\!+\!\rho\,\dvrg{\vec v}=0,  \label{rho}\\
&&\rho\der{\vec v}{t}\!+\!\nabla(p_\perp\!+\!\frac{B^2}{8\pi})\!
-\!\frac{1}{4\pi}(\vec{B}\cdot\!\nabla)\vec{B}\! =\!\rho{\vec
g}\!+\!(p_\perp\! -\!p_\parallel)\!\left[{\vec h}\dvrg{\vec
h}\!+\!({\vec h}\cdot\!\nabla){\vec h}\right]\!\!+\! {\vec h}({\vec
h}\cdot\!\nabla)(p_\perp\!-\!p_\parallel), \label{vel}\\
&&\der{}{t}\frac{p_\parallel B^2}{\rho^3}\!=\!
-\frac{B^2}{\rho^3}\!\left[B(\!{\vec h}\cdot\!\nabla\!)\!
\left(\!\frac{S_\parallel}{B}\right)\!+\!\frac{2S_\perp}{B}
 ({\vec h}\cdot\!\nabla)B\right], \ \label{ppar} \\
&&\der{}{t}\frac{p_\perp}{\rho B}= -\frac{B}{\rho}({\vec
h}\cdot\nabla)\left(\frac{S_\perp}{B^2}\right),
\label{pper}\\
&&\der{}{t}\frac{S_\parallel B^3}{\rho^4}\!=\! -j\frac{3p_\parallel
B^3} {\rho^4}({\vec h}\cdot\nabla)\!
\left(\frac{p_\parallel}{\rho}\right), \label{spar}\\
&&\der{}{t}\frac{S_\perp}{\rho^2}\!=\!-j\frac{p_\parallel} {\rho^2}
\left[({\vec h}\!\cdot\!\nabla)\!\left(\!\frac{p_\perp}{\rho}
\right)\!
+\!\frac{p_\perp}{\rho}\frac{p_\perp\!-\!p_\parallel}{p_\parallel
B}({\vec h}\!\cdot\!\nabla)B\right],\label{sper}\\
&&\der{\vec B}{t}+\vec{B}\dvrg{\vec
v}-(\vec{B}\cdot\nabla)\vec{v}=0, \label{ind} \\
&& \dvrg{\vec{B}}=0,\label{maks}
\end{eqnarray}
where $\nabla=\nabla_\parallel+\nabla_\perp, \nabla_\parallel={\vec
h}({\vec h}\cdot\nabla),$ and $
\der{}{t}=\pd{}{t}+(\vec{v}\cdot\nabla),
\,\vec{v}=\vec{v_\parallel}+
 \vec{v_\perp}, \,{\vec h}=\frac{\vec B}{B} $.
 Here $S_\|$ and $S_\bot$ are the heat fluxes along the magnetic
field by parallel and perpendicular thermal kinetic motions. If the
heat fluxes are neglected, $S_\bot=0$ and $S_\| = 0$, we obtain the
equations describing the laws of the change of longitudinal and
transverse thermal energy along the trajectories of the plasma (the
left-hand parts of Eqs. (\ref {ppar}--\ref {pper})). These so-called
``double-adiabatic'' parities and Eqs. (\ref {rho}), (\ref {vel}),
(\ref {ind}), and (\ref {maks}) form a closed system of equations,
the CGL (Chew-Goldberger-Low) equations, see the pioneering work by
Chew et al. (1956). However, the CGL-equations can result in
unsatisfactory heat flux evolution Eqs. (\ref {spar}--\ref {sper}).
This is because deducing the CGL equations the third moments of the
distribution function, hence the heat fluxes, have been lost without
any proof \cite{Chew,Baran}. The equations following from the
16--moments set in our case, Eqs. (\ref {rho}--\ref {maks}),
consider the heat fluxes, they are more complete, and the CGL
equations do not follow from these equations as a special case. One
should compare the final results in the limits $S_\bot \to 0$ and
$S_\|\to 0 $ with the results based on the CGL equations, deduced by
many authors \cite{Kato,Baran,Kuzn}.

To consider the CGL equations separately, we introduce on the
right-hand sides of the heat flux evaluation Eqs.
(\ref{spar}--\ref{sper}) the parameter $j$. To reach the exact CGL
equations we should take $S_{\|}=S_\bot=0$ and put $j=0$. In the
general non-CGL case $S_{\|}\ne 0$, $S_\bot\ne 0$, and $j\equiv 1$.

\section{Wave equations}

For simplicity we will now assume, that the basic initial
equilibrium state of the spatially non-limited plasma is
homogeneous, $g=0$, and the following quantities are constant: $v_0,
\rho_0, p_{\bot0}, p_{\|0}, B_0, S_{\bot 0}, \mathrm{and} \ S_{\|
0}$. Eqs. (\ref {rho}--\ref {maks}) will automatically satisfy such
an equilibrium state with non-zero initial heat fluxes. We will
consider small linear perturbations of all physical variables, e.g.
for pressure in the form $p=p_0+ p'(r, t) $. Let $p'(r, t) \sim \exp
i (\vec {k}\cdot\vec {r} - \omega t) $, where $ \omega = \omega_0 +
(\vec{v_0}\cdot \vec{k})$ is the wave frequency observed in the
moving frame of the fluid, and $k $ is the wave number of the
fluctuations. For the perturbations we obtain the equations
\begin{eqnarray}
&&\omega\rho' -\rho_0(\vec{k}\cdot\vec{v})=0,
\label{rho1}\\
&&\omega\rho_0\vec{v}
-\vec{k}\left(p_\bot'\!+\!\frac{\vec{B_0}\cdot\vec{B'}}{4\pi}\right)
\!+\! k_\parallel\frac{B_0}{4\pi}\vec{B'}\!-\!
\Delta\left[\vec{h_0}(\vec{k}\cdot\vec{h'})+k_\parallel\vec{h'}\right]
\!-\!k_\parallel\vec{h_0}\left(p_\|'-p_\bot'\right)\!=\!0, \label{vel1}\\
&&\omega\vec{B'} -\vec{B_0}(\vec{k}\cdot\vec{v})
+(\vec{k}\cdot\vec{B_0})\vec{v}=0, \\
&& (\vec{k}\cdot\vec{B'})=0, \label{far}\\
&& a_0\,\frac{p_\bot'}{p_{\bot0}}=a_1\frac{B'}{B_0} +
a_2\frac{\rho'}{\rho_0}, \
b_0\,\frac{p_\|'}{p_{\|0}}=b_1\frac{B'}{B_0} +
b_2\frac{\rho'}{\rho_0}. \label{per}
\end{eqnarray}
In deriving these equations, we have expressed the fluctuations of
the thermal fluxes as
\begin{eqnarray}
S_\bot'\!&\!=\!&\!j\frac{k_\parallel
p_{\|0}\,p_{\bot0}}{\omega\rho_0}\!\left[\frac{p_\bot'}{p_{\bot0}}\!
-\!\frac{\rho'}{\rho_0}\!-\!
\frac{\Delta}{p_{\|0}}\frac{B'}{B_0}\right]
\!+\!2S_{\bot0}\frac{\rho'}{\rho_0}\!, \label{Sper}\\
S_\|'\!&\!=\!&\!
j\!\frac{3p_{\|0}^2\,k_\parallel}{\rho_0\omega}\!\left(\frac{p_\|'}{p_{\|0}}
\!-\!\frac{\rho'}{\rho_0}\right)\!
-\!S_{\|0}\!\left(3\frac{B'}{B_0}\!-\!4\frac{\rho'}{\rho_0}\!\right)\!.
\label{Spar}
\end{eqnarray}
Here $\Delta=p_{\| 0}-p_{\bot 0}$, $\vec{h_0}=\vec{B_0}/B_0$,
$k_\parallel=(\vec{h_0}\cdot\vec{k})=k\cos\theta$. The indices $
\parallel $ and $ \perp $ correspond to the values of the parameters
along and across the magnetic field, respectively. Even if we insert
in Eqs. (\ref{Sper}--\ref{Spar}) $S_{\|0}=S_{\perp0}=0$, the
perturbations of these functions will never become zero:
$S'_{\|}\ne0$, $S'_{\perp}\ne0$. That means, using the 16--moments
equations we should get more reliable results on the wave properties
in an anisotropic plasma than with the CGL equations based on the
13--moments equations.

Strongly speaking, the initial heat fluxes are defined at the
kinetic level as the third moments of the particle distribution
function. In the presence of an external magnetic field the
components of this flux are defined by the steady solutions of the
kinetic equation. However, we should use here some appropriate
estimate as a parameter.

The initial collisionless heat flux functions  $S_{\|0}$ and
$S_{\perp0}$ should be estimated by taking the thermal energy
density of the electrons multiplied by the particle stream speed
along the magnetic field $u_0$. For example, for the solar wind
plasma we can write  $S_{\|0}\approx \frac{3}{2}
n_{\mathrm{e}}k_BT_\| u_0\,\delta = \frac{3}{4}\delta u_0 p_\|$.
Hollweg \cite{Hollweg74,Hollweg76} has given some estimates of the
correction parameter $\delta$ ($\alpha$ in his papers) assuming
various realistic shapes of the electron distribution function and
checking the results for agreement with space observations. $\delta$
depends on the magnetic field. In the range $B=0.1 - 100$\,G (1\,G =
$10^{-4}$\,T) we have $\delta\approx 4 - 0.1$. In the same way
$S_{\bot0}\approx \frac{3}{4}\delta u_0 p_\bot$.

Let us introduce dimensionless parameters (in the further text
indices `0' of physical parameters will be omitted for simplicity):
\be \alpha=\frac{p_\bot}{p_\|},\, \bar{\alpha}=1-\alpha, \,
c_\|^2=\frac{p_\|}{\rho},\, \beta=\frac{B^2}{4\pi
p_\|}=\frac{v_A^2}{c_\|^2}, \
\eta=\frac{c_\|\,k_\|}{\omega}=\frac{c_\|k}{\omega}\cos\theta,\ee
\be \bar{S_\|}=\frac{S_\|}{p_\|c_\|},\,
\bar{S_\bot}=\frac{S_\bot}{p_\bot c_\|}, \
\bar{S}=\alpha\bar{S_\bot}-2\bar{S_\|},\, l_1=\cos^2\theta,\,
l_2=\sin^2\theta. \ee Note that $\beta$ is defined here inversely
proportional to the often used plasma beta. Having in mind the
approximate estimates of the initial heat fluxes we may introduce
the dimensionless parameter $\gamma=(3/4)\delta u_0/c_\|$ by which
the heat fluxes are defined as $\bar{S_\|}=\bar{S_\bot}=\gamma$. By
means of these parameters the coefficients $a_{0,1,2} $ and
$b_{0,1,2} $ are defined as
\begin{eqnarray}
&& a_0=1-j\eta^2,\, a_1=1-2\gamma\eta-j\bar{\alpha}\eta^2, \,
a_2=1+2\gamma\eta-j\eta^2, \nonumber\\
&& b_0=1-3j\eta^2, \, b_1=2\gamma\eta(\alpha-2)-2,\,
b_2=3+4\gamma\eta-3j\eta^2. \label{ab}
\end{eqnarray}

With the above expressions and inserting Eqs. (\ref{rho1},
\ref{per}) we obtain from Eqs. (\ref{vel1}--\ref{far})
\begin{eqnarray}
&&\frac{1}{\eta}\frac{\vec v}{c_\|}-\alpha\frac{\vec
k}{k_\parallel}\left(\frac{a_1}{a_0}\frac{B'}{B}+
\frac{a_2}{a_0}\frac{({\vec k}\cdot\vec v)}{\omega}+\frac{(\vec
B\cdot\vec B')}{4\pi p_\bot} \right)+ \nonumber\\
&&+\frac{\vec B'}{B}(\beta-\bar{\alpha})-\frac{\vec B}{B}
\left[\left( \frac{b_1}{b_0} -\alpha\frac{a_1}{a_0}-2\bar{\alpha}
\right) \frac{B'}{B}+\right.
\left.\left(\frac{b_2}{b_0}-\alpha\frac{a_2}{a_0}\right)\frac{(\vec
k\cdot\vec v)}{\omega}\right]=0, \label{E1}\\
&& \frac{\vec B'}{B}-\frac{\vec B}{B}\frac{(\vec k\cdot\vec
v)}{\omega}+\frac{k_\parallel}{\omega}\,\vec{v}=0,\,\ \ (\vec
k\cdot\vec B')=0. \label{E2}
\end{eqnarray}
For the considered homogenous model we can place, without loss of
generality, both vectors of the unperturbed magnetic field $\vec B$
and the wave vector $\vec k$ into the same plane, say $x$--$z$. Then
the magnetic field perturbation vector $\vec B'$ is in the
perpendicular plane. Let the wave vector be along the $x$-axis. In
this geometry we have $\vec{k}=(k,0,0)$, $\vec{B}=(B_x, 0, B_z)$,
$\vec{B'}=(0, B'_y, B_z')$, and $\vec{v}=(v_x, v_y, v_z)$. Note that
$B_x=B\cos\theta$ and $B_z=B\sin\theta$. Taking the $x$, $z$ and $y$
components of the vector Eqs. (\ref{E1}--\ref{E2}) we get
\begin{eqnarray}
&&q_1\frac{v_x}{c_\|}-q_2\tan\theta\frac{B_z'}{B}=0, \,
\frac{1}{\eta}\frac{v_z}{c_\|}-q_3\tan\theta\frac{v_x}{c_\|}
+ q_4\frac{B_z'}{B}=0, \label{vz} \\
&&\frac{B_z'}{B}-\eta\tan\theta\frac{v_x}{c_\|}
+\eta\frac{v_z}{c_\|}=0, \,
 \frac{1}{\eta}\frac{v_y}{c_\|}+(\beta
-\bar\alpha)\frac{B_y'}{B}=0, \, \frac{B_y'}{B} +
\eta\frac{v_y}{c_\|}=0. \label{By}
\end{eqnarray}
Here $ q_4=\beta-\bar{\alpha}-l_2\,q_0$,
\begin{eqnarray}
  q_1 = \frac{1}{\eta}-\eta\frac{\alpha}{l_1}\frac{a_2}{a_0}-q_3,\,\,\,
  q_3=\eta\left(\frac{b_2}{b_0}
  -\alpha\frac{a_2}{a_0}  \right),\,\,\,
  q_2 = \alpha\frac{a_1}{a_0} + \beta+l_1 q_0,\,\,\,
   q_0=\frac{b_1}{b_0}-\alpha\frac{a_1}{a_0}-2\bar{\alpha}.&&\label{q14}
\end{eqnarray}
The last two $y$-Eqs. (\ref{By}) are separated from the others and
give the dispersion relation of fire hose modes
$\eta^2(\alpha+\beta-1)=1$ or
\begin{equation}\label{fh}
    \left(\frac{\omega}{k} \right)^2 =
     v_A^2\left(1-\frac{p_\|-p_\bot}{2p_\mathrm{m}} \right)\cos^2\theta.
\end{equation}
$p_\mathrm{m}$ is the magnetic pressure. The fire hose modes
(further the label $fh$ is used for these modes) are prototypes of
the Alfv\'en waves, and they become unstable if $\alpha+\beta < 1$
or if $p_\|> p_\bot + 2 p_\mathrm{m}$. The maximum of the
instability growing rate corresponds to the parallel propagation
case when $\cos^2\theta=1$. These modes are incompressible, and they
do not disturb the density of the plasma. The properties of these
modes remain unchanged including the heat flux evaluation Eqs.
(\ref{spar}--\ref{sper}).

The zero determinant of the first three $x-$ and $z-$equations of
Eqs. (\ref{vz}--\ref{By}) gives the dispersion relation for the
other modes:
\begin{equation}\label{dis0}
  l_2 q_2 \left(\frac{1}{\eta}-q_3 \right) - l_1 q_1\left(\frac{1}{\eta^2}-q_4 \right) =0.
\end{equation}

\section{CGL mirror wave instability}

To obtain the wave modes based on the CGL-invariants we should set
$\gamma=0$ and $j=0$ in the general dispersion Eq. (\ref{dis0}).
Then $a_0=a_1=a_2=b_0=1, b_1=-2, b_2=3$, and the dispersion equation
is
\begin{eqnarray}
 2\left(\frac{\omega}{k\ c_\|} \right)^2 = \alpha+\beta+2l_1+\alpha l_2 \pm \sqrt{A},
\ \ A= (\alpha+\beta+\alpha l_2 -4l_1)^2+ 4l_1 l_2 \alpha^2, && \label{cgl}
\end{eqnarray}
which has been obtained by many authors \cite{Kato,Baran}. Here the
fast and slow mirror mode waves correspond to the plus and minus
signs, respectively. Let the labels of these modes be $fm$ and $sm$.
In many aspects the properties of these well-known CGL modes are
similar to the usual MHD waves. For the parallel propagation case (
$l_1=1,\  l_2=0$) the squared phase velocities are
$V_{fm}^2=V_{fh}^2=\alpha+\beta-1$ and $V_{sm}^2=3$ if
$\alpha+\beta\ge 4$. In the opposite case $\alpha+\beta< 4$ we have
$V_{sm}^2=V_{fh}^2=\alpha+\beta-1$ and $V_{fm}^2=3$. In the
perpendicular propagation case ($l_1=0, \ l_2=1$)
$V_{sm}^2=V_{fh}^2=0$ and $V_{fm}^2=2\alpha+\beta$. Note that here
the phase velocities are normalized to $c_\|$.

However, there is a difference between the mirror modes and the MHD
magnetosonic waves: the relation $V_{sm}^2\le V_{fh}^2\le V_{fm}^2$
between the phase velocities is not always valid. In some parameter
ranges the slow modes may propagate faster than the Alfv\'enic fire
hose modes \cite{Kato,Hau93}. This behavior is opposite to the
isotropic MHD theory. The fast mirror modes are always stable
($V_{fm}^2>0$) as $A>0$. However, in some parameter ranges the slow
modes become unstable, $V_{sm}^2<0$. All of these properties can
easily be obtained from Eq. (\ref{cgl}). If both the fire hose and
the mirror instabilities arise at the same time the growing rate of
the first one is always greater, $\mathrm{Im}(V_{fh})\ge
\mathrm{Im}(V_{sm})$. The slow mirror instability condition is
\begin{equation}
l_2 \alpha^2 >3(\alpha+\beta+\alpha l_2-l_1).
\end{equation}
In the parallel propagation case ($l_1=1, \ l_2=0$) this passes to
the fire hose instability condition $\alpha+\beta<1$ or
$p_\|>p_\bot+2p_\mathrm{m}$. For the quasi-perpendicular modes
($l_1=0, \ l_2=1$) the mirror instability arises if $\alpha^2
>3(2\alpha+\beta)$ or $p_\bot^2/p_\|>6(p_\bot+p_\mathrm{m})$.
Compared to the kinetic theory the CGL theory gives the exact fire
hose instability criterion, but the mirror instability conditions
differs by a factor of 6 \cite{Tajiri,Has75,Lang02}.

\section{Instabilities with heat fluxes}

Let us now study our general dispersion Eq. (\ref{dis0}), obtained
without using the CGL invariants. In this equation $j\equiv 1$ and
generally $\gamma\ne 0$. This is a polynomial equation of 8\,th
order in the frequency of the fluctuations. For the parameter
$\eta=c_\| k_\|/\omega $ the dispersion equation can be written in
the form
\begin{eqnarray}
&& c_8\,\eta^8 + c_6\,\eta^6 + c_4\,\eta^4 + c_2\,\eta^2
 + c_0 + \gamma (c_7\eta^7 +
 c_5\,\eta^5 + c_3\,\eta^3)=0, \label{dis}
\end{eqnarray}
where for $l=l_1=\cos^2\theta$, $s=\alpha^2(1-l)$ ,
$$r=l-\beta-\alpha(2-l) ,\, c_8=3(2s+r) ,\, c_7=-4(3s+r), \,
c_5=4(s+r-l) ,$$
$$ c_6=2s(2\gamma^2-5) +3(l-3r) , \,
c_4=2s+7r-9l ,\,  c_3=4l , \, c_2=7l-r ,\, c_0=-l.$$
Here all the coefficients are real, consequently, all solutions are
real or conjugate complex. So, instead of the 4\,th order
biquadratic CGL dispersion equation we have deduced now the 8\,th
order Eq. (\ref{dis}) in the anisotropic MHD. With the initial heat
fluxes, $\gamma\ne 0$, odd nonzero coefficients $c_3, c_5, c_7$ will
result in wave propagation velocities depending on the propagation
direction with respect to the magnetic field. We can expect prograde
and retrograde wave modes. In the case of the CGL equations only two
mirror modes can arise, the phase velocities of which are equal to
each other in both directions with respect to the magnetic field.
Let us first consider the most important limiting and special cases
of Eq. (\ref{dis}) which can be solved analytically. It is useful to
represent Eq. (\ref{dis}) as follows:
\begin{eqnarray}
(\eta^2-1)(\mu\eta^2-l)(3\eta^4-6\eta^2+1-4\gamma\eta^3)-
2\alpha^2(1-l)\eta^4\wp=0 ,&& \label{dis1}
\end{eqnarray}
where $\wp=3\eta^4+(2\gamma^2-5)\eta^2+1+2\gamma\eta(1-3\eta^2)$
 and
$\mu=\alpha(2-l)+\beta-l$.

\subsection{Parallel propagation}

In the case of wave propagation along the magnetic field $l=1$,
$k_\|=k$, and the phase velocity normalized to the parallel sound
speed $c_\|$ is $V=1/\eta$. In this case the dispersion Eq.
(\ref{dis1}) becomes
\begin{eqnarray}
&&(\eta^2 -1)[\mu \eta^2 -1][3(\eta^2-
1+\sqrt{2/3})(\eta^2-1
 -\sqrt{2/3})-4\gamma \eta^3] = 0 , \label{dl0}
\end{eqnarray}
where $\mu=\alpha+\beta-1$. In contrast to the CGL case there appear
two additional modes which are connected with the heat fluxes. The
phase velocities of these two fast and slow thermal modes (the
corresponding labels are $ft$ and $st$) are between the modified
fast and slow mirror modes (corresponding labels are $fm$ and $sm$).
So in the phase diagrams these modes will be recognized by the
following relation between the real parts of the phase velocities:
\be\label{rel}
 V_{sm}^2 \le V_{st}^2 \le V_{ft}^2 \le V_{fm}^2.
\ee The place of fire hose mode velocity $V_{fh}^2$ in this relation
is arbitrary depending on $\alpha, \beta$, and $l$. In Eq.
(\ref{rel}) the relation `$\le$' means that a coincidence of the
mode branches is possible in a resonance interaction
 domains, where the instability is developing.

In Eq. (\ref{dl0}) the first two quadratic roots do not depend on
the heat flux parameter $\gamma$. That means, these modes are
symmetric with respect to the magnetic field direction. The last
four solutions do depend on $\gamma$. That means, waves running
along and against the magnetic field direction have different
velocities. For an easy classification of the solutions let us
consider the case $\gamma\to 0$. Here we should remember that
$\gamma=0$ means only that the initial fluxes are zero, but the
perturbed non-zero fluxes generated by the wave motions are
described by Eqs. (\ref{Sper}--\ref{Spar}). Now we have four
quadratic roots. It depends on the value of $\mu$ which root is
related to which mode. Let $\mu<0.55$. Then the root
$\eta^2=1-\sqrt{2/3}$ corresponds to the fast mirror mode with
$V_{fm}^2=3(1+\sqrt{2/3})\simeq 5.4$. The root $\eta^2=1$
corresponds to the fast thermal mode with $V_{ft}^2=1$. The other
root $\eta^2=1+\sqrt{2/3}$ corresponds to the slow thermal mode with
$V_{st}^2=3(1-\sqrt{2/3})\simeq 0.55$. The slowest mode is the slow
mirror mode corresponding to the root $\mu\eta^2=1$, for which
$V_{sm}^2=\alpha+\beta-1$. The fire hose modes coincide with the
slow mirror mode, $V_{fh}^2=V_{sm}^2$. However, with increasing
$\mu$ the fire hose mode can coincide with every of the obtained
modes. For instance, if $\mu>5.4$ then $V_{fh}^2=V_{fm}^2$. Such a
hierarchy between the phase velocities is similar to the CGL or MHD
theories. In the considered case the instability is possible if
$\mu<0$. This means only the slow mirror modes can be unstable.
Other wave modes propagating along the magnetic field are stable.

However, we have ignored  $\gamma$ in Eq. (\ref{dl0}). With
increasing $\gamma$ an additional instability is possible as well.
This can easily be verified considering a large $\gamma\gg 1$. The 4
solutions are: $\eta_1\approx 4\gamma/3$, $\eta_2\approx
(4\gamma)^{-1/3}$, $\eta_{3,4}\approx (4\gamma)^{-1/3}e^{\pm
2i\pi/3}$.  The phase velocities corresponding to these solutions
are: $V_1\approx 3/4\gamma\ll 1$, $V_2\approx (4\gamma)^{1/3}\gg 1$,
$V_{3,4}\approx (4\gamma)^{1/3}(-1 \pm i\sqrt{3})/2$. The two roots
of $\eta^2=1$ give $V_{5,6}=\pm 1$. In the fire hose instability
parameter range ($\mu<0$) the roots of the last equation
$\mu\eta^2=1$ correspond to $V_{7,8}=\pm i\sqrt{1-\alpha-\beta}$.
Now we apply the relation Eq. (\ref{rel}) to the waves propagating
along ($\mathrm{Re}(V)>0$) and against ($\mathrm{Re}(V)<0$) the
magnetic field separately. For the positive phase velocities we
have: $V_{sm}^+=\mathrm{Re}(V_7)=0$, $V_{st}^+=V_1\ll 1$,
$V_{ft}^+=V_5=1$, $V_{fm}^+=V_2\gg 1$. For the negative phase
velocities we get: $V_{sm}^-=\mathrm{Re}(V_8)=0$, $V_{st}^-=V_6=-1$,
$V_{ft}^-=\mathrm{Re}(V_3)$, $V_{fm}^-=\mathrm{Re}(V_4)$. The
squares of these negative phase velocities also satisfy Eq.
(\ref{rel}). From here we conclude the role of the nonzero initial
heat fluxes: \\ i) Retrograde thermal modes are faster than prograde
thermal modes, $|V_{st}^-|\gg V_{st}^+$ and $|V_{ft}^-|\gg
V_{ft}^+$. For the fast mirror waves we have the opposite case:
prograde modes are twice faster than the retrograde modes,
$V_{fm}^+=2|V_{fm}^-|$. Slow mirror modes remain symmetric. If we
let the parameter $\gamma$ go back to zero, all the discrepancies
between the same modes disappear. \\ ii) Instabilities have a
resonance origin. Instabilities grow when two (maybe even more)
phase velocities coincide. The aperiodic fire hose instability
develops when the velocities of the slow mirror modes are zero,
$V_{sm}^+=V_{sm}^-=0$. The periodic instability (non-zero real
frequency) appears, when the retrograde fast thermal and mirror
modes are in resonance, $V_{ft}^-=V_{fm}^-\ne0$. Depending on the
values of $\gamma$ the grow rate of the thermal mirror instability
may be greater than the fire hose growing rate.

\subsection{Quasi-perpendicular propagation}

In the case of quasi-perpendicular propagation ($l\to 0$) inserting
$\eta=l^{1/2}X$ into Eq. (\ref{dis}) results in
\begin{eqnarray}
 l^3c_8\,X^8 + l^2c_6\,X^6 + lc_4\,X^4 + c_2\,X^2 -1 +
 \,l^{3/2} \gamma\, (lc_7\,X^4 + c_5\,X^2 + 4)X^3 = 0, && \label{dper}
\end{eqnarray}
where the coefficients $c_{2-8}$ are the same as in Eq. (\ref{dis})
if we insert there $l=0$ and the phase velocity $V=1/X$. The first
two solutions of this equation for $l=0$ are
$X^2=1/c_2=1/(2\alpha+\beta)$. These are equal to the stable CGL
fast mirror modes with the phase velocity
$V_{1,2}=\pm\sqrt{2\alpha+\beta}$. To find the other 6 solutions of
Eq. (\ref{dper}) we introduce the new variable $X=Y/\sqrt{l}$. Then
the equation for $Y$ is
\begin{equation}\label{eqy}
 c_8\,Y^6\!+\!c_6\,Y^4\!+\!c_4\,Y^2\!+\!c_2\!+\!\gamma Y^3 (c_7\,Y^2\!+\!c_5) = 0.
\end{equation}
Let us consider the case $\gamma\to 0$. For $Z=Y^2$ we have now
\begin{equation}\label{eqz}
 c_8\,Z^3 + c_6\,Z^2 + c_4\,Z + c_2 = 0.
\end{equation}
Remind that the remaining 6 phase velocities are expressed by three
exact solutions of this cubic equation,
$V_{3-8}=\pm\sqrt{l/Z_{1,2,3}}$. The known analytical solutions of
Eq. (\ref{eqz}) are still cumbersome. But for some parameter limits
we can give these solution in simple expressions. Let
$\sigma=(2\alpha+\beta)/(2\alpha^2)$. Then Eq. (\ref{eqz}) is
\begin{equation}\label{eqz1}
 3(1-\sigma)\,Z^3\!+\!(9\sigma-5)\,Z^2\!+\!(1-7\sigma)\,Z\!+\!\sigma = 0.
\end{equation}
We will consider three limit cases: $\sigma\approx 1$, $\sigma\ll
1$,
 and $\sigma\gg 1$.\\
i) Expansions of the solutions around $\sigma=1$ give
$Z_{1,2}\approx (3\pm\sqrt{5})/4$ and $Z_3\approx 4/3(\sigma-1)$.
Correspondingly the phase velocities are $V_{3-6}\approx
\pm\sqrt{l(3\pm\sqrt{5})}$ and $V_{7-8}\approx
\pm\sqrt{3l(\sigma-1)/4}$. The first 4 solutions are stable as
$V_{3-6}^2>0$, but the last ones may become unstable $V_{7-8}^2<0$
if $\sigma<1$ is obeyed. This corresponds to the mirror instability
criterion well known from kinetic plasma physics \cite{veden58,
Kich60,Has75}, $2\alpha+\beta<2\alpha^2$ or
\begin{equation}\label{mir}
    \frac{p_\bot^2}{p_\|} >p_\bot + p_\mathrm{m}.
\end{equation}
Contrary to the CGL case the factor 6 does not appear on the r.h.s.
of this condition.\\
Expanding the dispersion equation around small $l\ll 1$ we obtain
 the growing rate of the mirror instability
\begin{equation}\label{gromir}
\frac{\omega}{kc_\|}\approx \pm i
\frac{3l}{2}\sqrt{1-\sigma-l\frac{(\alpha-1)(2\alpha+1)}{2\alpha^2}}
.
\end{equation}
The maximum of this growing rate is
$\sqrt{6}\alpha(1-\sigma)/[4\sqrt{(\alpha-1)(2\alpha+1)}]$
 which corresponds to a critical angle of
$l_c=\alpha^2(1-\sigma)/[(\alpha-1)(2\alpha+1)]\le 1$. Qualitatively
this result for the fluid approximation is in good agreement with
similar kinetic results, such as for the guiding center theory
\cite{Kuls}, the low frequency analytical limit of kinetic
turbulence \cite{veden58}, and the more exact numerical results for
different kinds of particle distribution functions \cite{Gedal02}.

 ii) The limit case $\sigma\ll 1$ corresponds to
$\alpha\gg 1$ or $p_\bot\gg p_\|$ (hot particle across the magnetic
field limit). In this case $V_{3-6}\approx
\pm\sqrt{l(5\pm\sqrt{13})/2}$ and $V_{7-8}\approx \pm
i\sqrt{l/\sigma}$. Again the first 4 mode solutions are stable, the
last two modes become unstable. The growing rate of these modes may
be smaller or larger depending on
the ratio $l/\sigma$. \\
iii) The limit $\sigma\gg 1$ is also interesting. This limit
corresponds to $\alpha\ll 1$ ($p_\bot\ll p_\|$ -- hot particles
along the magnetic field) or $\beta\gg 1$ -- strong magnetic field.
In this case we have only stable modes: $V_{3-6}\approx
\pm\sqrt{l(3\pm\sqrt{6})}$ and $V_{7-8}\approx \pm \sqrt{l}$.

If we apply Eq. (\ref{rel}) to the deduced phase velocities we
obtain for $\sigma\approx 1$ that $V_{fm}^2=V_{1,2}^2$,
$V_{ft,st}^2=V_{3-6}^2$, and $V_{sm}^2=V_{7,8}^2$. These definitions
are the same as those for the case $\sigma\gg 1$, but for $\sigma\ll
1$ this definition is strongly depending on the ratio of the two
small parameters $l/\sigma$.

For large $\gamma\gg 1$ it can easily be shown from the asymptotical
solutions of Eq. (\ref{eqy}) that the mirror instability condition
(\ref{mir}) remains unchanged.

\subsection{Oblique propagation}

For the oblique propagation case ($l\ne 1$ and $l\ne 0$) it becomes
more difficult to analyze Eq. (\ref{dis}). However, there are some
important limit cases for which simple analytical solutions are
possible.

\subsubsection{The case $\alpha\ll 1$}
This case is similar to the parallel propagation case. Here Eq.
(\ref{dis1}) is reduced to Eq. (\ref{dl0}) with the difference that
the parameter $\eta$ contains $l$ and the phase velocity
$V^2=l/\eta^2$. Besides, instead of the equation $\mu\eta^2=1$ we
have $\mu\eta^2=l$ with $\mu=\beta-l$. That means slow mirror waves
become unstable at oblique propagation: $V_{7,8}=\pm\sqrt{\beta-l}$.
Instability appears if $\beta<l$.

\subsubsection{The case $\alpha\gg 1$}
In this case  $V=\sqrt{l}/\eta$ and the 4 solutions of Eq.
(\ref{dis}) are
\begin{equation}\label{ag1}
 V^2\approx \frac{\alpha}{2}[1+l_2\pm\sqrt{(1+l_2)^2+8ll_2}],
\end{equation}
where $l_2=1-l$. The `+' and `-' solutions are fast and slow mirror
modes, respectively. These modes do not depend on $\gamma$, and they
are symmetric with respect to the propagation direction. The fast
modes are stable, but unstable are the slow modes with large growing
rate. The instability develops aperiodically.

The other 4 solutions for the thermal modes are described by the
equation
\begin{equation}\label{ag2}
 3\eta^4+(2\gamma^2-5)\eta^2+1+2\gamma\eta(1-3\eta^2)=0.
\end{equation}
For $\gamma\ll 1$ the thermal modes are stable, $V^2\approx
l(5\pm\sqrt{13})/2>0$. With increasing $\gamma$ instability appears.
In the limit $\gamma\gg 1$ the prograde modes remain stable  with
$V\approx \sqrt{l}(3\pm\sqrt{3})/2\gamma$, but the retrograde modes
become strongly unstable with
\begin{equation}\label{ag3}
    V\approx \gamma\sqrt{l}(-1\pm i).
\end{equation}
The real parts of the phase velocities of the fast and slow thermal
modes coincide, and a periodical thermal instability develops.

\subsubsection{The case $\beta\gg 1$}
A strong magnetic field $\beta=2p_m/p_\|\gg 1$ (reverse plasma beta)
is an often used special case. In this case the solutions of Eq.
(\ref{dis}) are also simple. The fast and slow mirror modes are
symmetrical and stable, $V_{fm}^2\approx\beta$ and $V_{sm}^2\approx
l$. If the heat flux parameter $\gamma\ll 1$ then both the fast and
the slow thermal modes are symmetric and stable too, $V^2\approx
3\pm\sqrt{6}$. However, with increasing $\gamma$ there appears an
asymmetry and an instability of the thermal modes. For $\gamma\gg 1$
the prograde modes are asymmetric but stable, $V_{ft}\approx
\sqrt{l}(4\gamma)^{1/3}$ and $V_{st}\approx 3\sqrt{l}/(4\gamma)$.
The phase velocities of the two retrograde modes become equal and
they are unstable: $V \approx \sqrt{l}(4\gamma)^{1/3}(-1\pm
i\sqrt{3})/2$.

\subsubsection{The case $\gamma\gg 1$}
This special case of a large heat flux parameter, but moderate
parameters for the rest is also of some theoretical interest.
Assuming $l\ne 0$ and $l\ne 1$  we obtain that the two fast mirror
modes are asymmetrical and stable: $V^\pm_{fm}\approx\pm
l^{1/2}[2\gamma(g\pm l)/l]^{1/3}$, where
$g=(l^2+\alpha^2l\,l_2)^{1/2}$, and the signs `$\pm$' correspond to
the propagation direction on magnetic field. For the two slow mirror
modes we have
\begin{equation}\label{gg1}
 V^\pm_{sm}\approx \frac{\sqrt{l}}{2\gamma\nu_1}\!\left(\!3\nu_1
 -\nu_2\pm\sqrt{3\nu_1^2+
\nu_2^2-3\nu_1\nu_2}\right).
\end{equation}
Here $\nu_1=\alpha^2(1-l)$, $\nu_2=\alpha(2-l)+\beta-l$. As the
expression under the square root is positive these modes are stable.
The 4 thermal modes become unstable. For the prograde two thermal
modes
\begin{equation}\label{gg2}
V^+_{ft,st}\approx l^{1/2}[2\gamma(g - l)/l]^{1/3}(1\pm i\sqrt{3})/2
\end{equation}
and for the two retrograde thermal modes we have
\begin{equation}\label{gg3}
V^-_{ft,st}\approx l^{1/2}[2\gamma(g + l)/l]^{1/3}(-1\pm
i\sqrt{3})/2.
\end{equation}

\begin{figure*}
\centering
\hfill\includegraphics[width=0.4\textwidth]{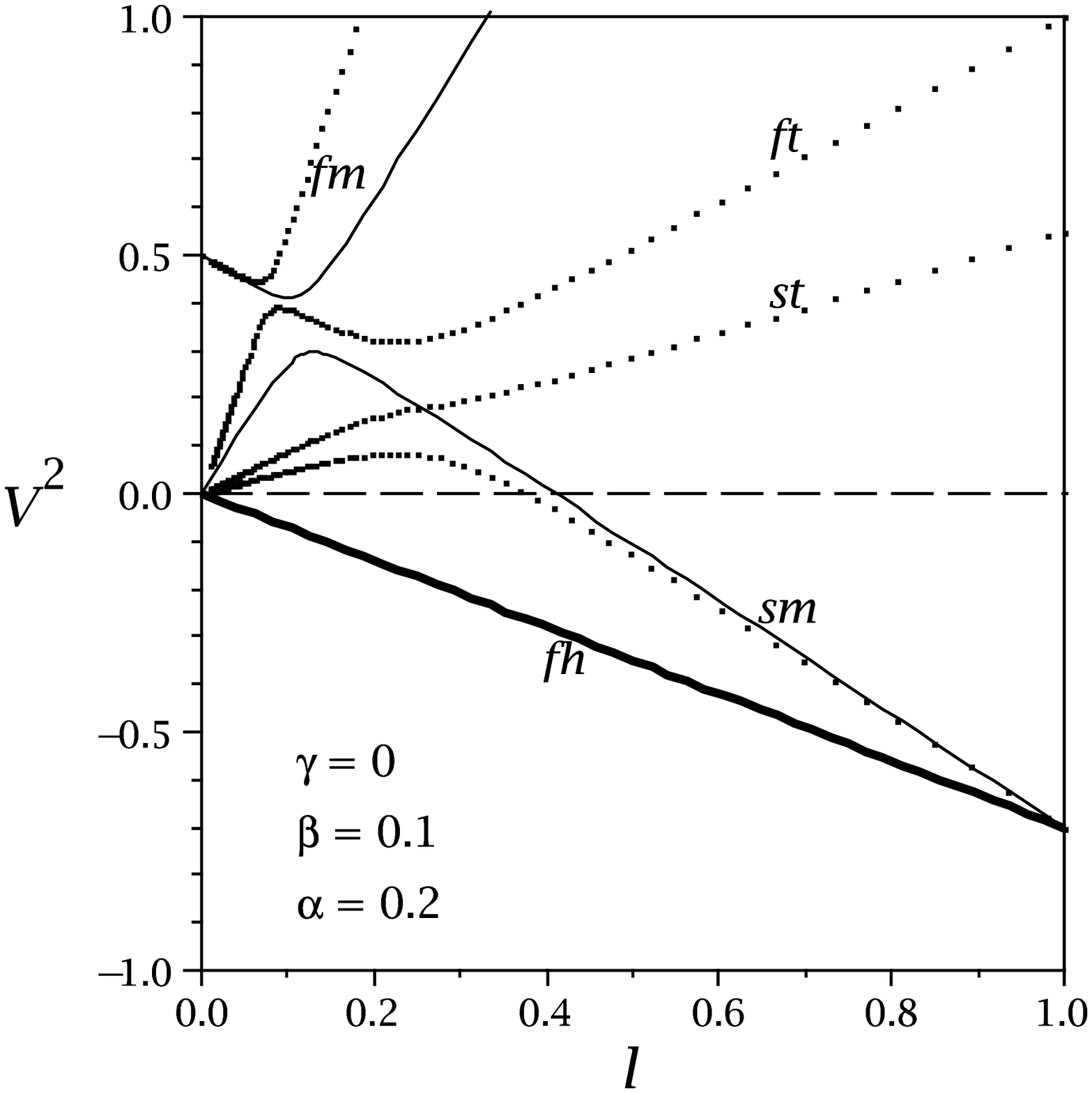}%
\hfill\includegraphics[width=0.4\textwidth]{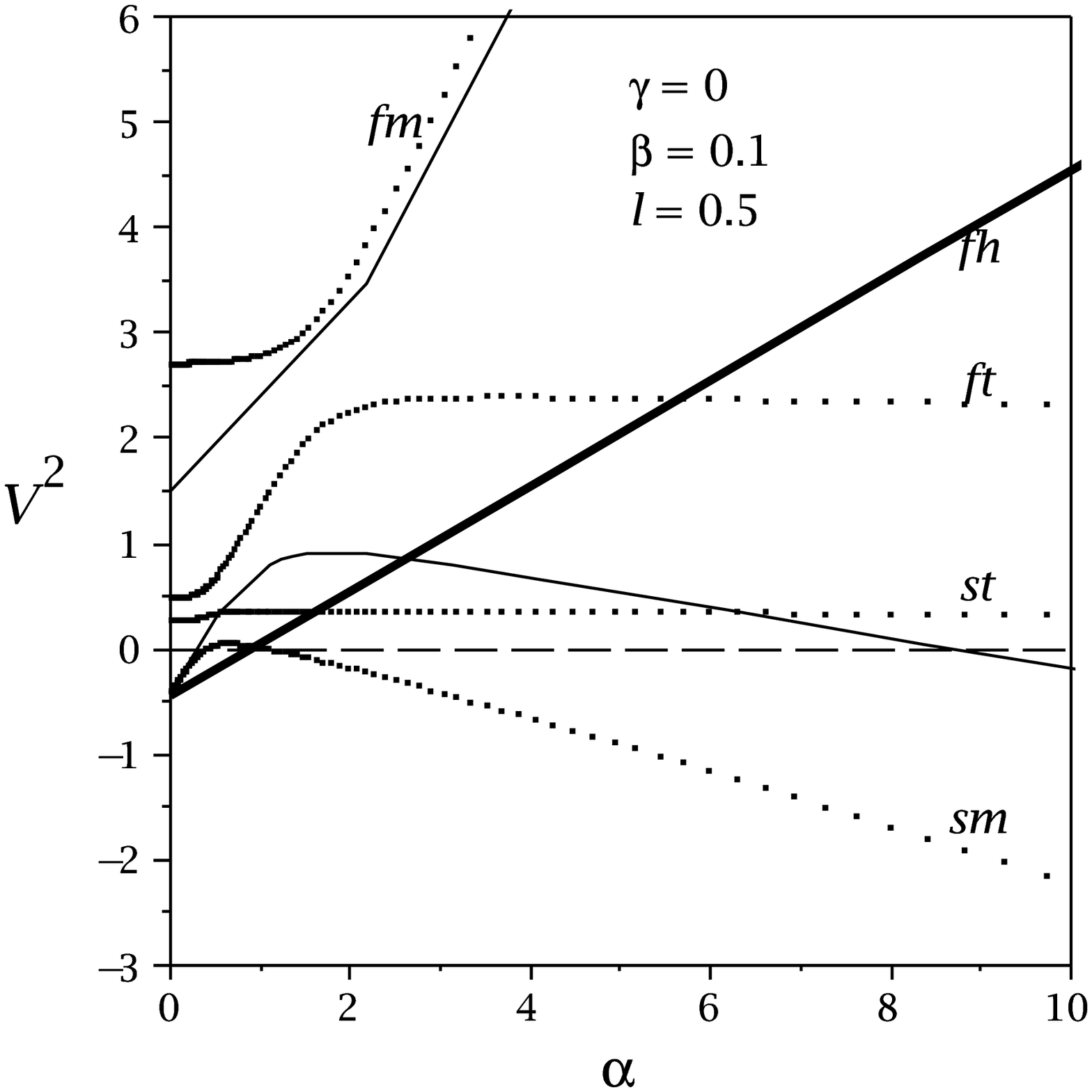}\hspace*{\fill}
\caption{The normalized phase velocity squared as function of the
wave propagation angle $l=\cos^2\theta$ (left-hand picture) and of
the anisotropy parameter $\alpha=p_\bot/p_\|$ (right-hand picture)
for absent initial heat fluxes, $\gamma=0$. In the area below the
dashed line where $V^2<0$ the modes become unstable for
$\mathrm{Re}(\omega)=0$. The dotted lines are the 4 wave mode
solutions of the dispersion Eq. (\ref{dis}). The 2 thin solid curves
are the CGL fast and slow mirror modes. The thick solid line is the
classic fire hose mode. The labels at the curves correspond to the
mode classification: $fm$ - fast mirror, $sm$ - slow mirror, $ft$ -
fast thermal, $st$ - slow thermal, and $fh$ - fire hose modes. }
\label{gam0}
\end{figure*}

\begin{figure*}
\centering
\hfill\includegraphics[width=0.4\textwidth]{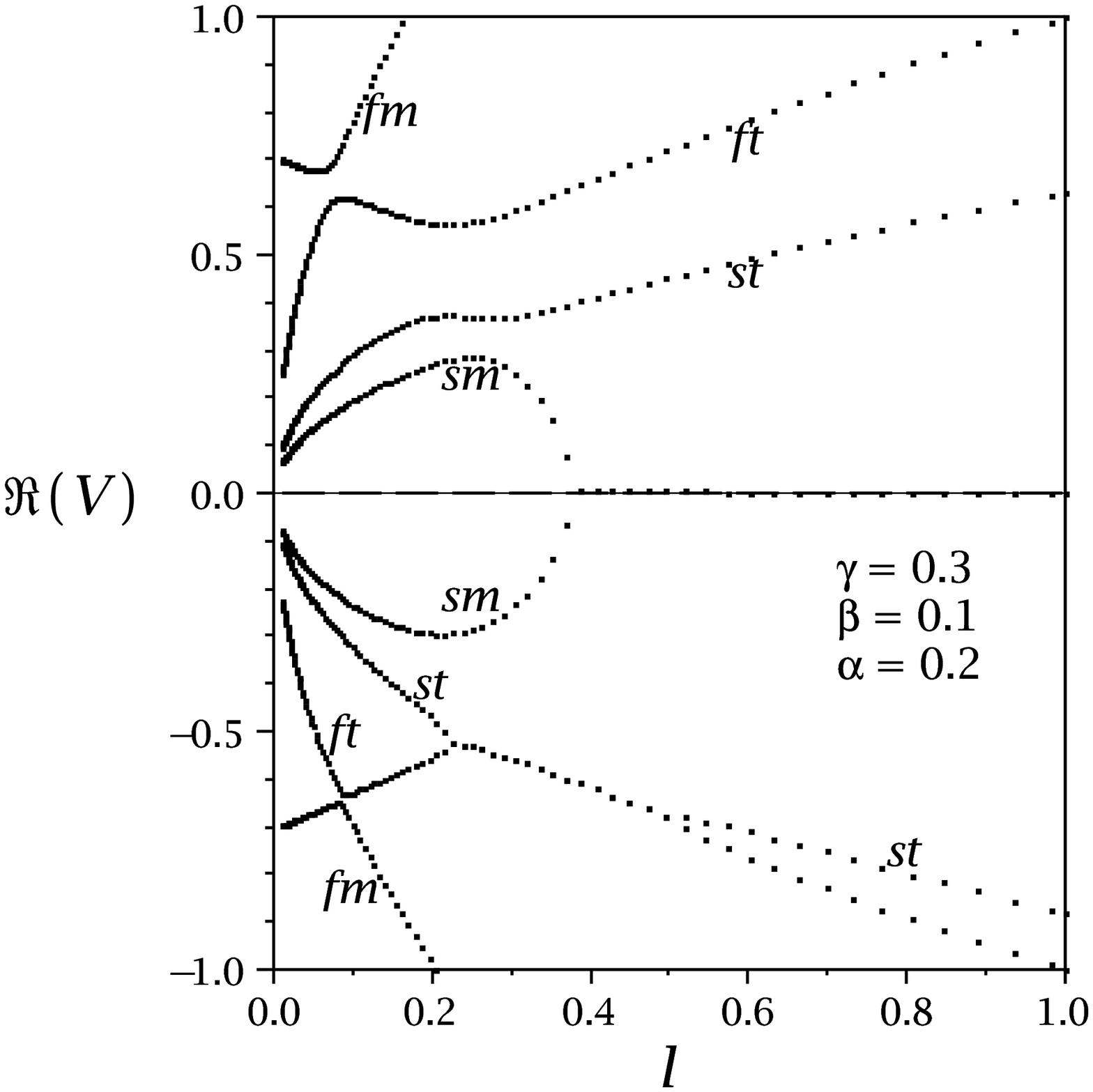}%
\hfill\includegraphics[width=0.4\textwidth]{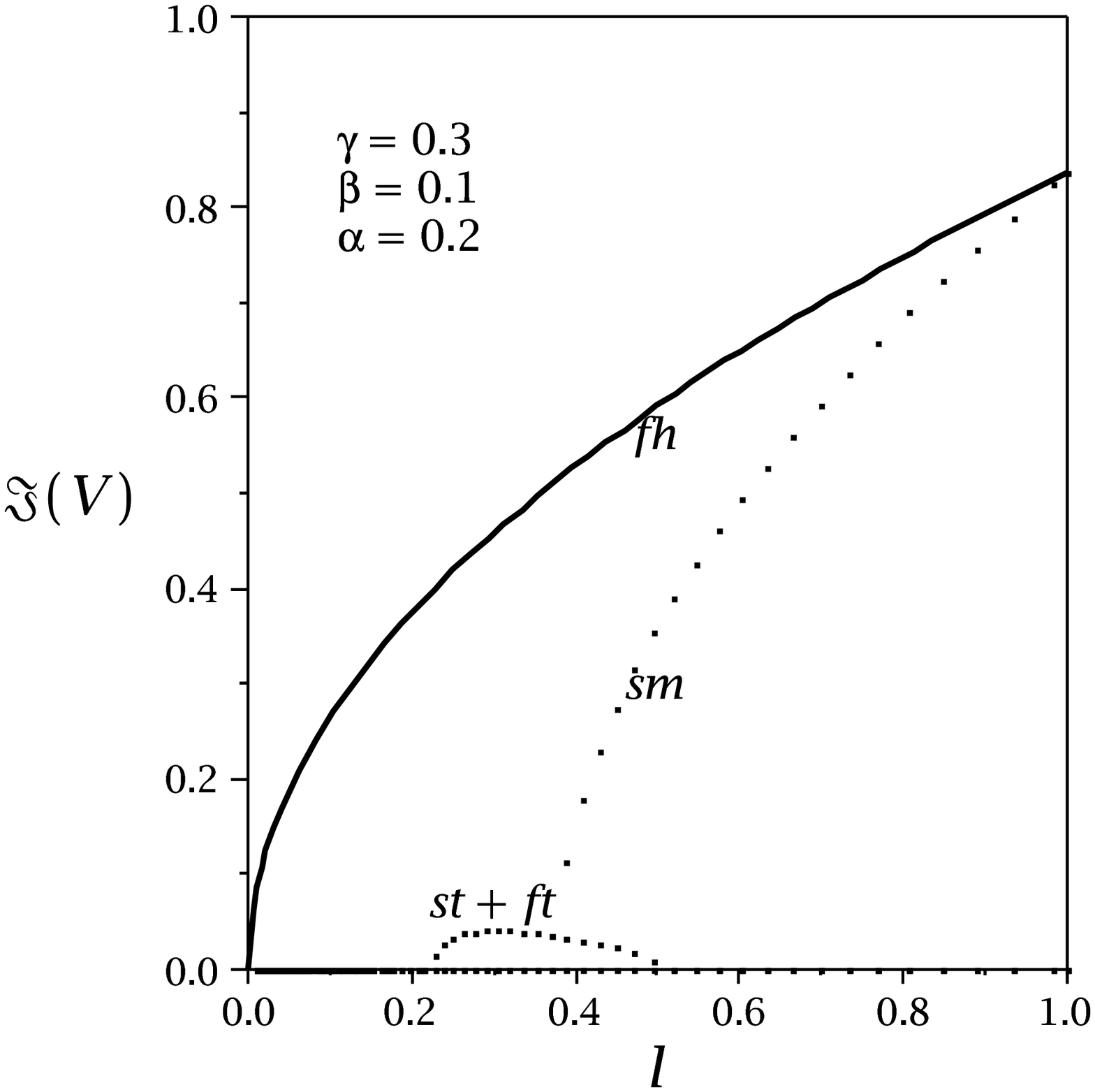}\hspace*{\fill}
\caption{The same as the left-hand picture of Fig. \ref{gam0}, but
with non-zero heat fluxes, $\gamma=0.3$. The left-hand picture is
for the phase velocities $\mathrm{Re}(\omega)/kc_\|$, the right-hand
picture for the growing instability rates
$\mathrm{Im}(\omega)/kc_\|$ in dependence on the propagation angle
parameter $l$. Instability arises when the phase velocities of the
different modes in the left-hand picture coincide. In the right-hand
picture the shown fire hose and slow mode instabilities are
aperiodical ($\mathrm{Re}(\omega)=0$), while the thermal instability
is periodical ($\mathrm{Re}(\omega)\ne 0$). } \label{gam03}
\end{figure*}

\begin{figure*}
\centering
\hfill\includegraphics[width=0.4\textwidth]{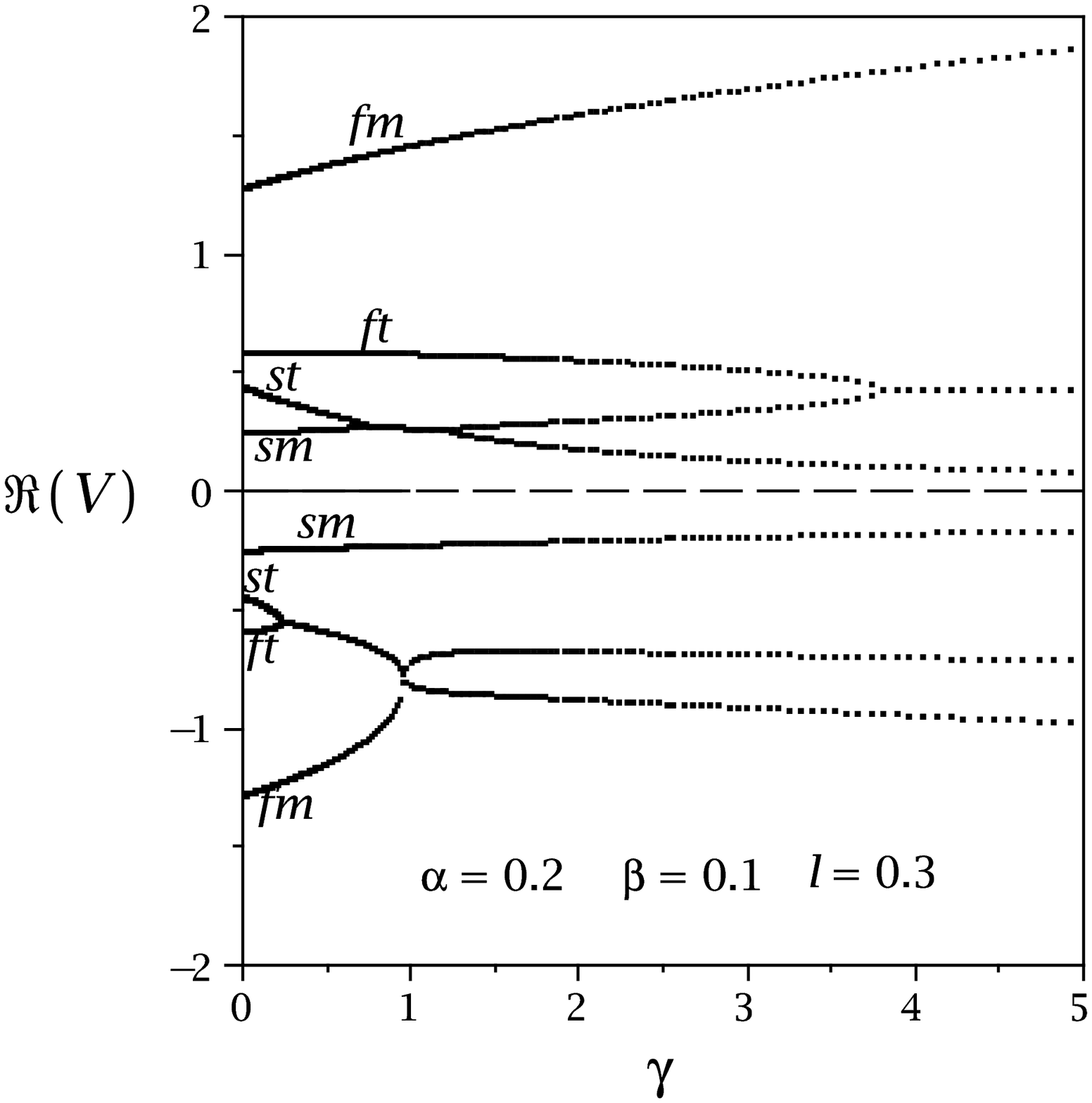}%
\hfill\includegraphics[width=0.4\textwidth]{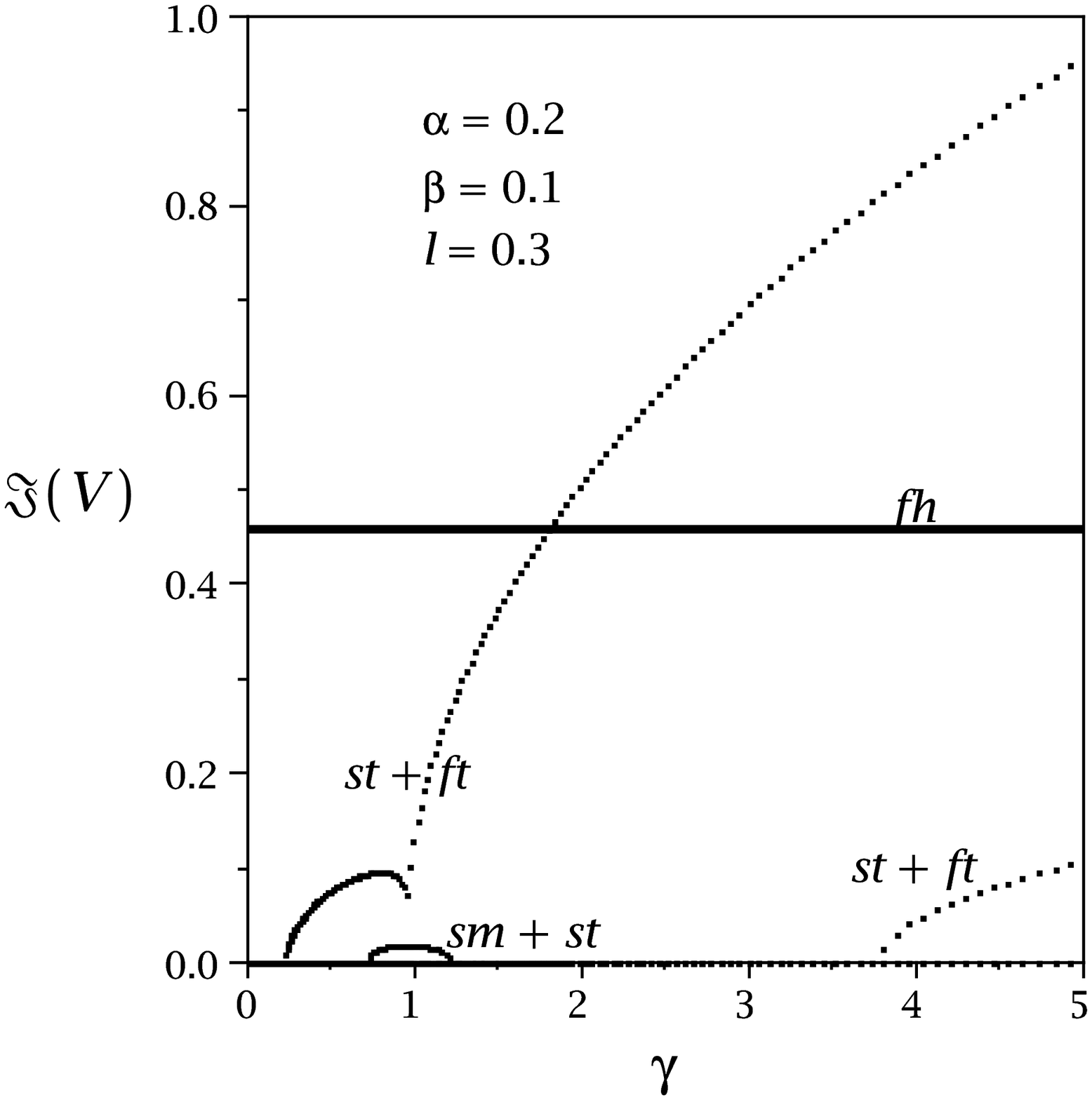}\hspace*{\fill}
\caption{Phase velocities (left-hand picture) and growing
instability rates (right-hand picture) as functions of the heat flux
parameter $\gamma$. Prograde modes ($\mathrm{Re}(V)>0$) become
unstable in two ranges of $\gamma$: $\gamma\approx 1$ where slow
mirror and slow thermal modes are in resonance (the label $sm+st$ in
the right-hand picture) and $\gamma>3.8$ where two thermal modes are
in resonance (the label $st+ft$ in the lower right corner of the
right-hand picture). The larger instability rate with label $st+ft$
is due to the resonance of the two retrograde thermal modes. The
fire hose mode ($fh$) does not depend on $\gamma$ .} \label{gam}
\end{figure*}

\subsubsection{Some numerical examples}
The coefficients of the dispersion Eq. (\ref{dis}) depend on four
parameters. In realistic space plasmas the values of these
parameters cover wide ranges: $\alpha>0$, $\beta>0$, $\gamma>0$, and
$0\le l \le 1$. Here we cannot illustrate all interesting parameter
ranges, but we will show some typical examples for high plasma-beta
($=2/\beta$), which cannot be considered asymptotically. The first
two pictures shown in Fig. \ref{gam0} are normalized squared phase
velocities $V^2=\omega^2/k^2c_\|^2$ in the symmetrical case of
$\gamma=0$. In these pictures our results (the 4 dotted curves) are
compared with the well known CGL modes (the 2 thin curves). It is
seen that both CGL modes are strongly modified, especially the slow
mirror mode. The left-hand picture demonstrates the fire hose
instability development when $\alpha+\beta<1$. With a decrease of
the propagation angle both unstable slow modes tend to the fire hose
mode and they are equal in the parallel propagation. For all angles
the fire hose instability is dominant compared to the slow mode
mirror instability. The right-hand picture shows the fire hose
instability disappearing with increasing $\alpha$. Here we see the
strong differences between the slow modes in the CGL and the new
theory. The modified slow modes are highly unstable for $\alpha>1$.
If the initial heat fluxes are absent ($\gamma=0$) both additional
thermal modes are stable.

The second set of two pictures shown in Fig. 2 are similar to those
on the left-hand side of Fig. \ref{gam0} (growing of fire hose
instability), but now with non-zero initial heat fluxes, $\gamma\ne
0$. The slight deviation of $\gamma$ from zero has a strong
influence on the retrograde thermal modes. In some range of the
parameter $l$ two fast and slow thermal modes resonantly interact.
In this region the phase velocities are equal and a periodical
thermal instability develops. For the considered parameter values
and small $\gamma$ the other modes are only slightly changed. The
aperiodical fire hose and the slow mode mirror instabilities also
arise when the prograde and retrograde phase velocities become equal
to zero. The found new thermal instability arises strongly for
oblique propagation. For the considered parameters this range
corresponds to $0.2<l<0.5$, where the instability rate has a
maximum. As seen from the right-hand picture the fire hose
instability is still dominant.

In the last two pictures in Fig. \ref{gam} we show how the new
thermal instability is changed with increasing $\gamma$. We choose
$l=0.3$ where in Fig. \ref{gam03} the thermal instability has its
maximum. With $\gamma$ the growing rate of these retrograde thermal
modes increases very sharply, and it becomes larger than the growing
rate of the fire hose instability. To compare it with the fire hose
instability in the right-hand picture the level of the growing rate
of the fire hose instability is shown too. Here appear also the
instabilities of the prograde modes, but with smaller growing rates.
Slow thermal modes interact at first with the slow mirror modes
close to $\gamma\approx 1$, then for higher $\gamma>4$ they are in
resonance with the fast thermal modes.

\begin{figure}
\noindent\centering{\includegraphics[width=0.4\textwidth]{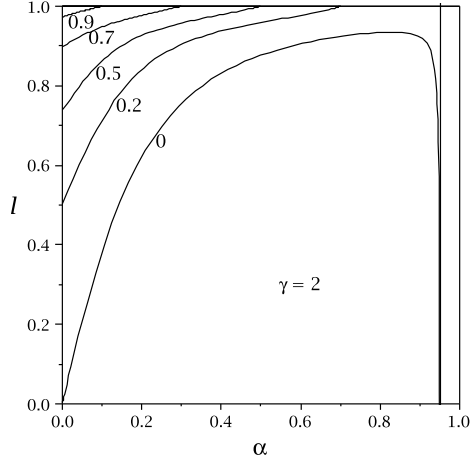}}
\caption{Compressible fire hose instability domains in dependence on
$\alpha$, $l$, and $\beta$ (the numbers at the curves) for given
$\gamma=2$.} \label{f2l2}
\end{figure}

\section{Compressible fire hose instability}

The classical incompressible fire hose instability arises if
$\eta^2=-a$ and $a=1/(1-\alpha-\beta)>1$ are obeyed, see Eq.
(\ref{fh}). Near the threshold of this instability $a\gg 1$. This is
an aperiodical instability, that means $\mathrm{Re}(\omega)=0$. Here
we show that the dispersion equation for the compressible modes Eq.
(\ref{dis}) can reach a similar solution at the threshold of the
fire hose instability but with a small non-zero real frequency. For
this aim we will search the solution of Eq. (\ref{dis}) in the form
\begin{equation}\label{fh2}
    \eta^2=-a\,(1+i\,\epsilon),
\end{equation}
where $\epsilon$ is a new unknown which is real. For simplicity let
$|\epsilon|\ll 1$ and $\eta\approx \mp ia^{1/2}(1+i\epsilon/2)$.
Then Eq. (\ref{dis}) is split into two equations which correspond to
its real and imaginary parts. The first one defines the parameter
$\epsilon$:
\begin{equation}\label{eps1}
\epsilon \approx \pm \gamma a^{3/2}
 \frac{c_7a^2-c_5a+c_3}{4c_8a^4-3c_6a^3+2c_4a^2-c_2a}.
\end{equation}
Near the instability threshold where $a\gg 1$ we get a simple
relation for $\epsilon$:
\begin{equation}\label{eps2}
\epsilon \approx \mp \frac{\gamma}{3\sqrt{a}}\,
\frac{1+\alpha-3\alpha^2}{(1-\alpha)(2\alpha+1)}.
\end{equation}
Here we should remember that $\alpha < 1$. The second equation is
\begin{eqnarray}
c_8a^4-c_6a^3+c_4a^2-c_2a+c_0=
\frac{\gamma^2a^3}{2}\,\frac{(c_7a^2\!-\!c_5a\!+\!c_3)
(7c_7a^2\!-\!5c_5a\!+\!3c_3)}{4c_8a^4-3c_6a^3+2c_4a^2-c_2a}. &&
\label{l12}
\end{eqnarray}
This is a quadratic equation for $l=\cos^2\theta$:
$d_1l^2+d_2l+d_3=0$. The coefficients $d_{1, 2, 3}$ are real
functions of the parameters $\alpha, \beta$, and $\gamma$. The
cumbersome expressions of these coefficients can be obtained easily
from Eq. (\ref{l12}).  In Fig. \ref{f2l2} we show these solutions in
the range $0\le l\le 1$ for $\gamma=2$. Only in the fire hose
instability parameter values, when $\alpha<1$ and $\beta<1$, we get
$l$ in the range $0\le l\le 1$. The $l(\alpha, \beta)$ picture is
not strongly changed for values $0\le\gamma\le 2$. The
quasi-parallel propagation of $l\approx 1$ is easily described by
Eq. (\ref{dis1}). For the found $l(\alpha, \beta, \gamma)$ the
compressible periodical fire hose instability can develop in this
way as described by Eq. (\ref{fh2}) with $\epsilon(l)$ defined by
Eq. (\ref{eps1}). $\epsilon$ strongly depends on $\gamma$. For
absent heat fluxes, $\gamma=0$, we have $\epsilon=0$ and the found
oblique instability disappears. The complex phase velocity in this
instability is defined as
\begin{equation}\label{f2V}
V\approx \pm(\epsilon/2 +i)\sqrt{l/a}.
\end{equation}
Formally this formula  is the same as that for the incompressible
fire hose modes if we set $\gamma=0$. The difference between both
modes is that the first one can develop for any $l$ in the given
$\alpha$ and $\beta$ ranges, but the second fire hose instability
develops only for one angle of propagation defined by $l(\alpha,
\beta, \gamma)$. Taking into account Eq. (\ref{f2V}) that
$\mathrm{Re}(V)<0$. So the second fire hose instability is the
result of the resonantly coupling of the retrograde thermal modes.

\begin{figure*}
\centering
\hfill\includegraphics[width=0.4\textwidth]{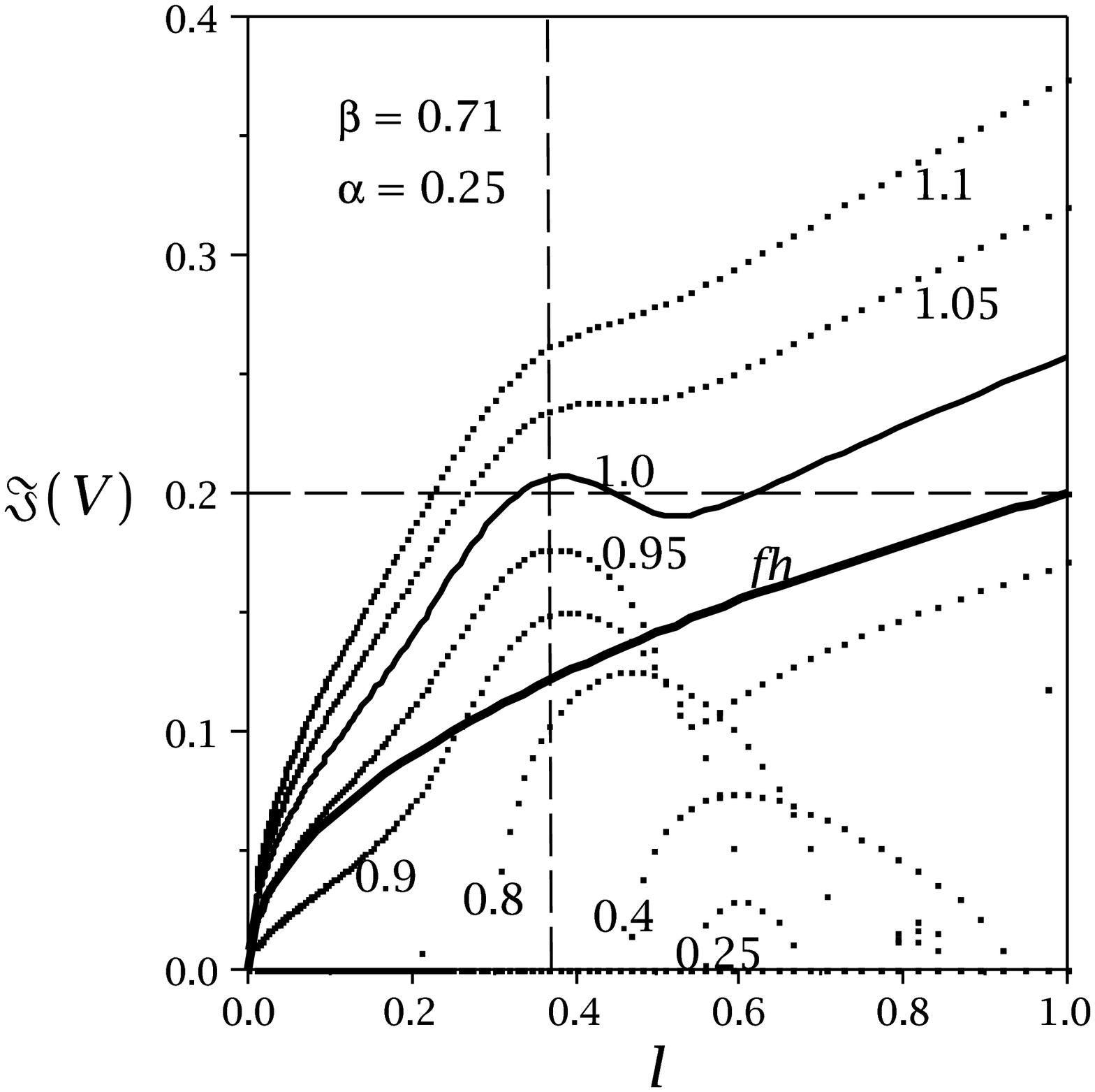}%
\hfill\includegraphics[width=0.4\textwidth]{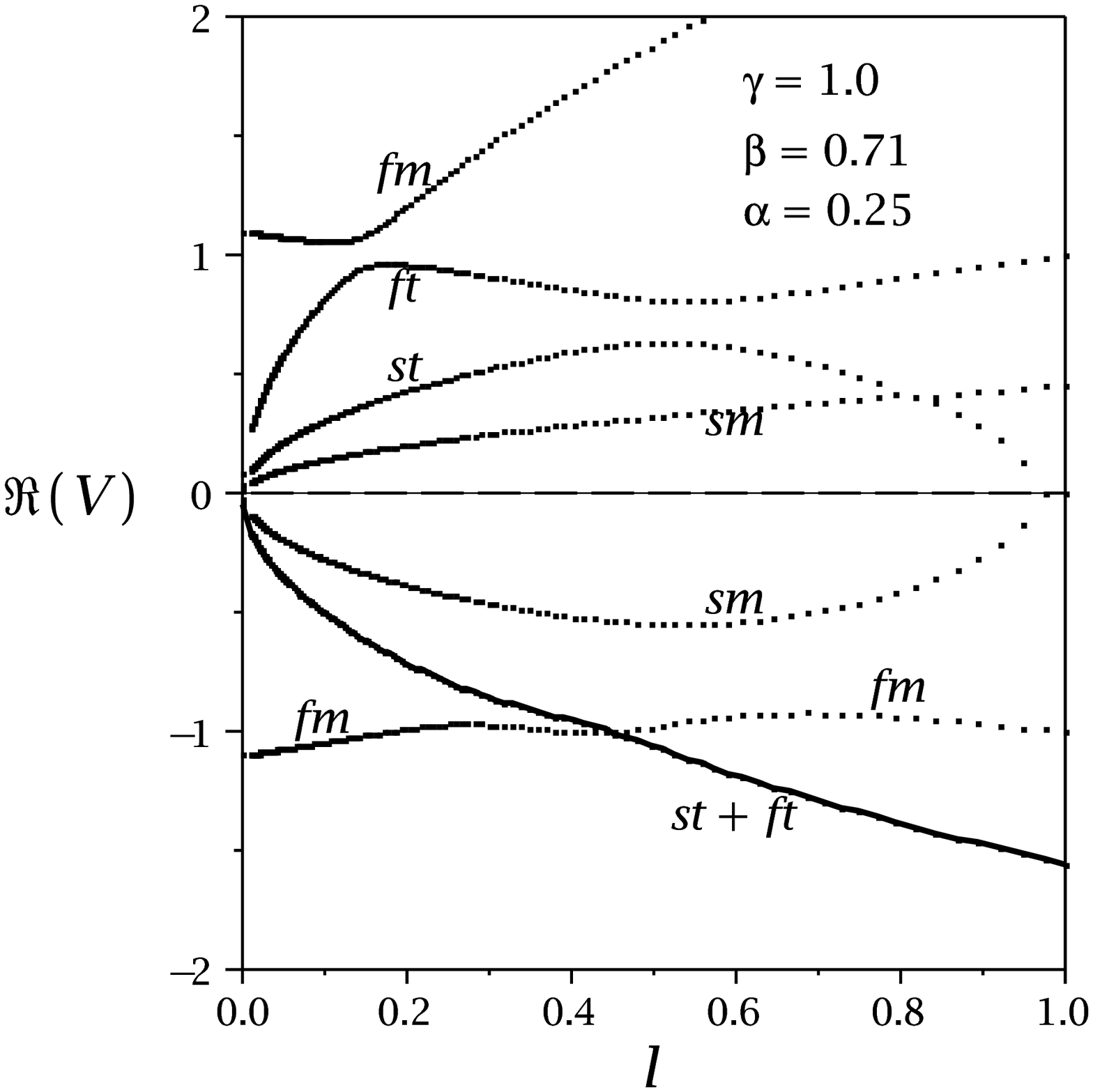}\hspace*{\fill}
\caption{Left-hand picture: compressible fire hose instability
growing rate as function of $l$ and $\gamma$ (numbers at the
curves). The thick solid curve is the growing rate of the
incompressible fire hose instability, and the horizontal dashed line
is the level of its maximum for parallel propagation ($l=1$). The
vertical dashed line corresponds to the angle $\theta=53^o$
($l=0.36$) when the growing rates of the two kinetic fire hose
instabilities become comparable to each other. For fluid
instabilities this condition is obeyed for $\gamma\approx 1$.
Right-hand picture: phase velocities for the case of $\gamma=1$.  }
\label{hell}
\end{figure*}

We think that the found second instability is analogous to the
earlier found kinetic oblique fire hose instability
\cite{Hellinger00}. Both the incompressible and the compressible
instability growing rates become comparable to each other near the
threshold of the fire hose instability. At the given high proton
plasma beta (=2.8) the oblique compressible fire hose instability
rate at the angle $\theta=53^o$ reaches (or even slightly exceeds)
the maximum of the incompressible fire hose instability rate in
parallel propagation. We got here a possibility to test this proton
anisotropy instabilities in our fluid approximation. We take the
same parameters. In our definitions $\beta=0.71$, $l=0.362$, and
$\alpha=0.25$ (close to the threshold). However, we have one free
parameter -- the parameter of the heat fluxes $\gamma$. Of course,
in the kinetic consideration such a parameter does not exist. So
this free parameter has to be varied to fit the results of the
kinetic study. For this calculations the condition $\epsilon\ll 1$
is not used in Eq. (\ref{fh2}). The results are shown in Fig.
\ref{hell}. It is seen from the left-hand picture that an analogous
situation is reached for the fluid fire hose instabilities in the
range $\gamma \la 1$. The right-hand picture shows the phase
velocities of the 8 modes for $\gamma=1$. The retrograde fast
thermal modes are coupled at low $l$ with the slow thermal modes,
and at higher values of $l$ these modes are in resonance with the
fast mirror modes. For the value $l\sim 0.4$ of interest for us all
3 retrograde modes are in resonance. So the interaction of the 3
retrograde modes results in the appearance of the second kind of the
fire hose instability.

\begin{figure*}
\centering
\hfill\includegraphics[width=0.4\textwidth]{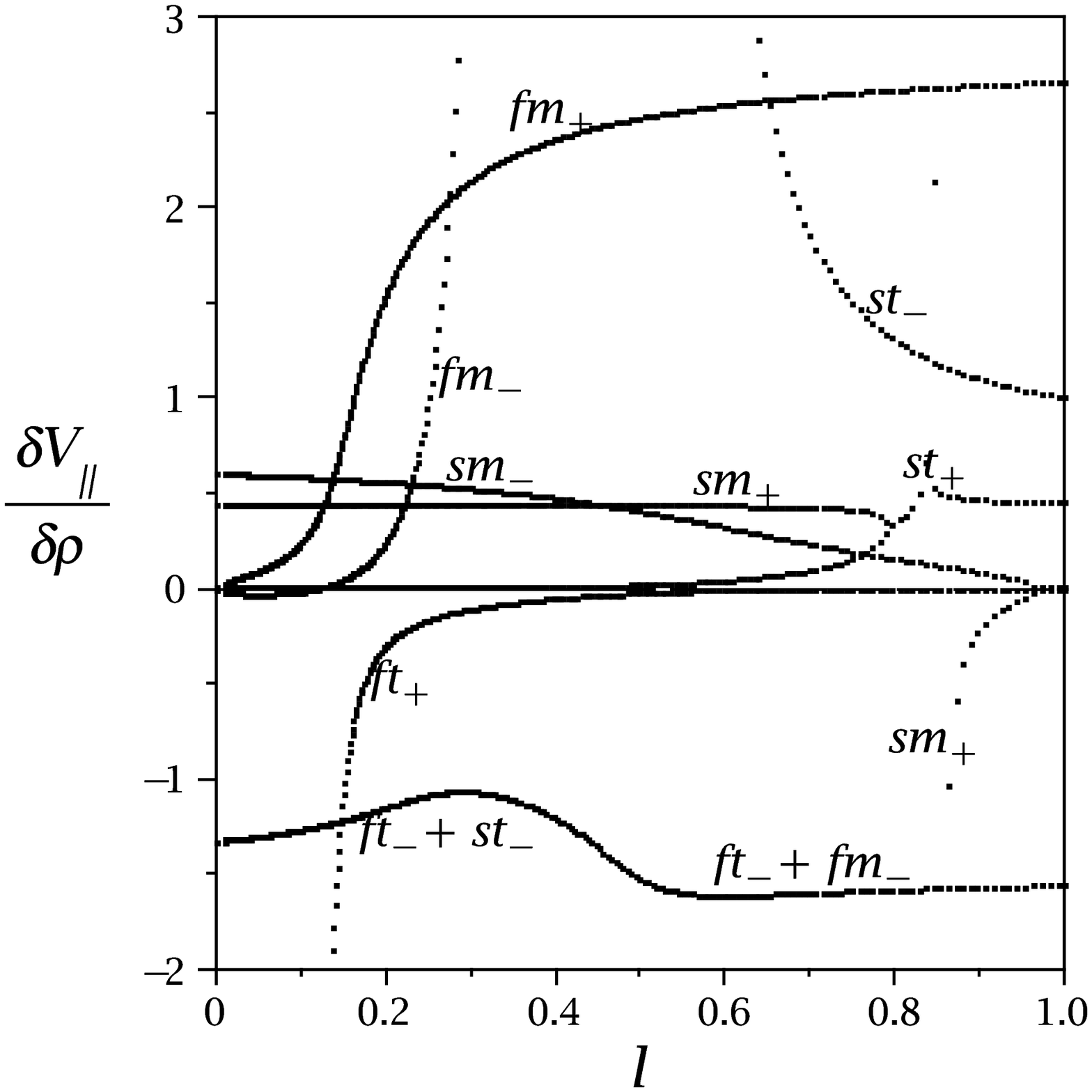}%
\hfill\includegraphics[width=0.4\textwidth]{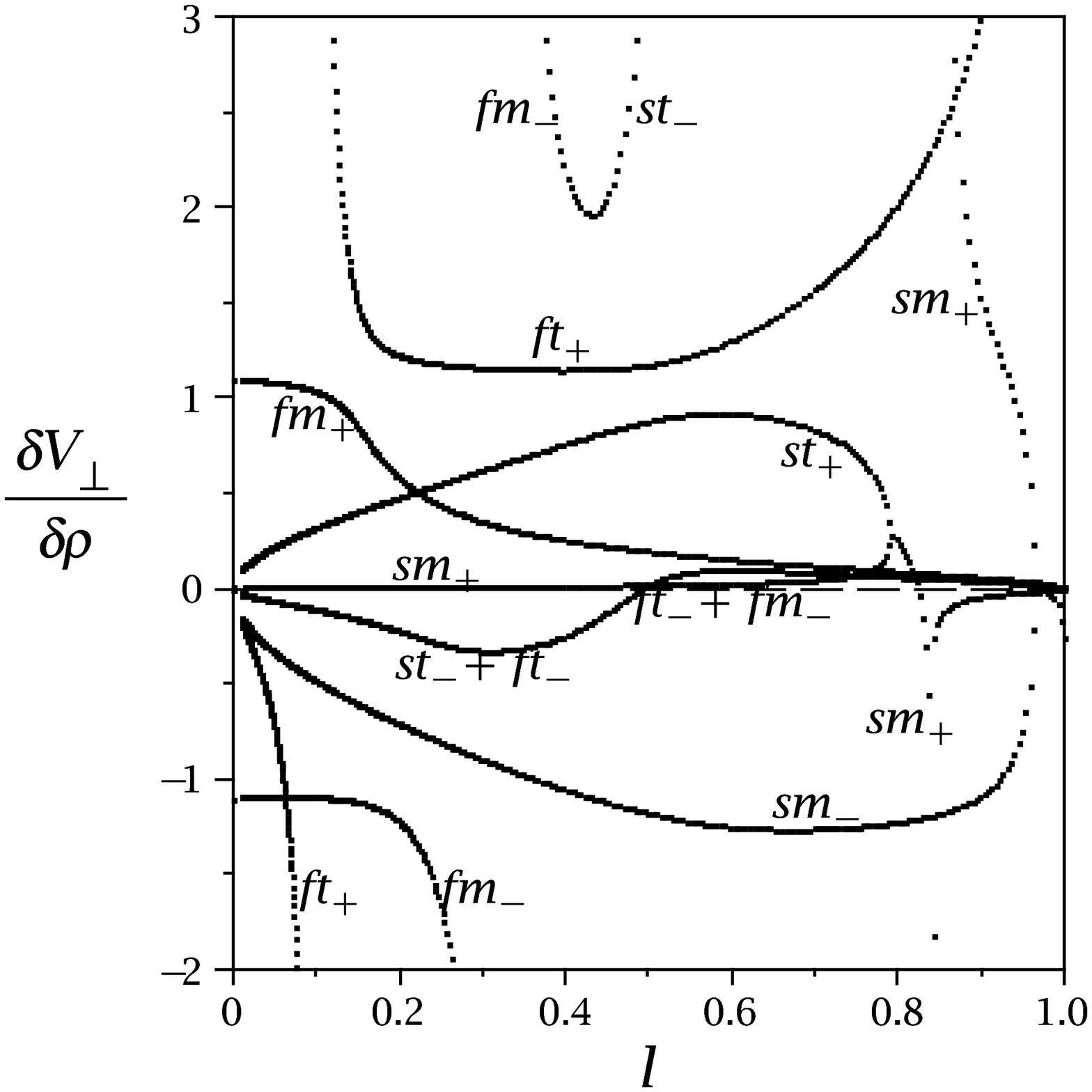}\hspace*{\fill}
\caption{Examples of fluid velocity and density perturbation
amplitude relations in dependence on $l=\cos^2\theta$ for given
other parameters $\alpha=0.25$, $\beta=0.71$, and $\gamma=1$. Here
$\delta\rho=\mathrm{Re}(\rho'/\rho_0)$, $\delta
V_\|=\mathrm{Re}(v_\|/c_\|)$ and $\delta
V_\bot=\mathrm{Re}(v_\bot/c_\|)$. The last quantities are normalized
fluid velocity components along and across the magnetic field. The
labels at the curves correspond to the wave classification marks,
and plus and minus at the indices of the labels correspond to
prograde and retrograde wave modes, respectively. } \label{rhopp}
\end{figure*}

\section{Mass density fluctuations}

The wave modes discussed above are compressible, they produce
density perturbations. Here we shall write down further formulae for
these perturbations which can be used in practice for identifying
the wave motions.
\begin{eqnarray}
&&\alpha l_2\frac{a_2}{a_0}\frac{\rho'}{\rho_0}\!=\!
\left[l_1\!\left(\frac{1}{\eta^2}\!-\!\alpha\!-\!\beta\!+\!1\right)\!-\!
l_2\!\left(\beta\!+\!\alpha\frac{a_1}{a_0}\right)\!\right]\!\frac{B'}{B_0}, \nonumber\\
&&\frac{v_\|}{c_\|} = \frac{1}{\eta}\left(1-l_2\frac{1-\eta\,
q_3}{1-\eta^2 q_4}\right) \frac{\rho'}{\rho_0}, \
\frac{v_\bot}{c_\|} = \frac{\sqrt{l_1l_2}}{\eta}\,\frac{1-\eta\,
q_3}{1-\eta^2 q_4}\,\frac{\rho'}{\rho_0}. \label{rhoB}
\end{eqnarray}
$v_\|$ and $v_\bot$ are fluid velocity components along and across
the magnetic field, $l_1=\cos^2\theta$, $l_2=\sin^2\theta$, and
other parameters are defined by Eqs. (\ref{ab}) and (\ref{q14}). The
inverse parallel component of the complex phase velocity $\eta=k_\|
c_\|/\omega$ is defined as a solution of the dispersion relation Eq.
(\ref{dis}). These formulae allow us to restore the full components
of the fluid velocity and the magnetic field perturbation amplitudes
if any components of velocity and density perturbations are known
from observations.

Fig. \ref{rhopp} shows an example of the ratios of the parallel and
perpendicular velocity amplitudes to the density perturbation
amplitude. This case is equal to that on the right-hand picture of
Fig. \ref{hell} where $\alpha=0.25$, $\beta=0.71$, and $\gamma=1$ is
considered. Such a pictures gives direct information on the
polarization of every type of wave modes. For instance, it is seen
that all wave modes are almost longitudinally polarized  at parallel
propagation, $l=1$. It follows from the first relation of Eq.
(\ref{rhoB}) (the picture is not shown here) that the usual
isotropic MHD relationship of $\delta\rho \delta B<0$ for the slow
MHD mode and  $\delta\rho \delta B>0$ for the fast MHD mode is not
valid here. The sign of $\delta\rho \delta B$ can change in
dependence on the parameters values.

\section{Discussion and Conclusions}

\subsection{Model equations}
Compared with a full kinetic model the fluid description of a plasma
has the mathematical advantage of a smaller number of dimensions.
Moreover, many observed dynamic phenomena in space plasmas are
large-scale structures --- already averaged over both temporal and
spatial scales. This suggests to find ways for a description of a
plasma as a fluid. It is easily possible in the case of a
collision-dominated plasma which is described by a Maxwellian
distribution function. In this case the usual isotropic MHD
equations are received. The situation becomes much more complicated
for a smaller frequency of the collisions between the particles of a
hot magnetized plasma (such as space plasmas in most cases). Due to
the magnetic field the collisionless plasma becomes anisotropic with
respect of its local direction. To describe in this case the plasma
as a fluid the transport model equations are deduced from the
kinetic equations for the moments of the distribution function.
These moments are such quantities as plasma mass density, fluid
speed of the plasma, anisotropic thermal pressure, anisotropic
thermal flux, etc. Basically, the number of these moments is
infinite, and the equations of these moments are coupled among each
other. However, if the conditions $ \omega / \Omega_{B} \ll 1$ and
$k\, r_{B} \ll 1$ are satisfied ($ \omega $ -- frequency of
perturbations, $ \Omega_{B} $ -- gyration frequency, $k $ -- wave
number, $r_{B} $ -- gyration radius), it is possible to break off
the chain of these equations, if some additional conditions related
to the given exact analytical form of the particle distribution
function are fulfilled. This method is called the standard method of
MHD ordering of the kinetic equations, and the resulting new
equations are called ``transport equations''. However, the
application of this ordering method has some subtleties -- it is not
trivial \cite{Ramos03}. Depending on these and on additionally
chosen conditions, the obtained transport equations can be
different. By including higher order moments it is possible to
increase the accuracy of the transport equations. With increasing
order of the ordering the accuracy of the model equations also
increases, but they become more complicated for the analysis. Even
though these equations can never be complete without supposing
additional conditions, they describe well such phenomena as
Alfv\'{e}nic and acoustic (electronic and ionic) waves, they can
include such an important kinetic effect as Landau damping. However,
depending on the order of ordering such kinetic effects as
drift-waves and other micro-instabilities can be lost. Besides, to
deduce the transport equations an exact analytical type of the
function of particle distribution with anisotropic temperatures,
e.g. bi-Maxwellian  or $k$-distributions, is required. That means,
the fine structure of the realistic distribution functions and the
related microphysics are ignored.

Among these model equations the CGL equation are most simple with
respect to the included moments and the order of ordering. The heat
flux tensor in these equations is ignored at all. Strictly speaking,
i.e. the phase speed of the perturbations should be much larger than
the thermal velocity of the particles. Such condition are met for
Alfv\'{e}nic modes, but for acoustic modes this condition is
impracticable. The inclusion of a heat flux tensor to the equations
was carried by many authors \cite{Barak,Snyder,Mahajan}. Results of
our studies in the present paper (types of instabilities, conditions
of their existence, thresholds and values of growing rates of
instabilities, comparisons of these with the corresponding results
for mirror and ion-acoustic instabilities in the low-frequency
kinetic approach) let us come to the conclusion that the used
equations derived by Ramos \cite{Ramos03} are more correct.

To sum it up it can be said that we have used more complete fluid
transport equations describing the macroscopic behavior of a
magnetized anisotropic collisionless plasma. In particular, we
include heat fluxes and their evolution which, contrary to the
CGL-MHD model, are a basic feature and cannot be ignored. We
consider only two parallel heat fluxes corresponding to parallel and
perpendicular thermal motions of particles. Perpendicular heat
fluxes have been neglected.

\subsection{Mirror instability}
In strongly magnetized and weakly collisional turbulent plasmas the
anisotropy of the pressure develops in a spontaneous way
\cite{Schekoch,Sharma}. In a high-beta plasma that triggers a number
of instabilities, above all firehose and mirror
\cite{chandra58,parker0,veden58,barnes66,Hasegawa69}. The nonlinear
evolution of the instabilities should have the tendency to
compensate on the average the pressure anisotropies generated by the
turbulence. Thus the development of the instability of the modes
further changes the distribution function, tending to make it more
isotropic, or to strengthen the magnetic structurization. If $p_\bot
> p_ \| $, there appear two kinds of electromagnetic instability:
ion-cyclotron instability at frequencies $ \omega < \Omega _ {Bi} $
($ \Omega _ {Bi} $ -- ion cyclotron frequency) and magneto-mirror
instability at a very low frequencies, $ \omega\approx 0$. Because
there exist numerous observations of low-frequency turbulence in
magneto-active plasmas, for instance in magnetosheaths, in solar
wind, and in cometary comas (see references in \cite{Gedal01}) the
mirror instability was studied theoretically in detail. It was shown
that in ionic high-beta plasma the mirror modes become unstable if
an anisotropy indicator $p_\bot / p_ \| - 1 $ exceeds some critical
value \cite{chandra58,veden58}. This instability causing a local
deformation of the magnetic field makes the plasma spatially
inhomogeneous.  It occurs because a part of the particles captured
in ``weak mirror traps'' subdivides the distribution of particles
into passing and trapped species \cite{Kivel}.

The fluid analogy of the kinetic mirror instability has a similar
simple description \cite{Hasegawa69,Thompson}. Basically, the
instability occurs because at low frequencies the changes of the
perpendicular pressure of the plasma and of the magnetic field occur
in opposite phases. Really, as follows from Eq. (\ref{per}), for
$\omega\to 0$ neglecting small density perturbations, $p_\bot' \sim
- p_\bot (\frac{p_\bot}{p_ \|} - 1) \frac {B'}{B}$. Hence, in those
places of the plasma where $p_\bot >  p_ \| $ a decrease of the
plasma pressure increases locally the intensity and the pressure of
the magnetic field. Meeting the condition $ \frac {p_\bot}{p_ \|} -1
> \frac{p_{\mathrm m}}{p_\bot}$ the force (caused by the total pressure) in
perpendicular direction decreases. Thus, the increase in intensity
of the magnetic field locally reduces the total pressure which, in
its turn, pushes together the magnetic field lines even more, i.e.
leads to a further growth of the magnetic field. It causes
instability. From the invariance of the first magnetic moment of the
plasma follows that the energy of the particles will simultaneously
grow in the direction perpendicular to the magnetic field. From the
conservation of total energy follows that the parallel energy should
decrease accordingly. Under these conditions (when the magnetic
moment and the total energy are conserved) the plasma will naturally
flow from an area with high magnetic field intensity to an area with
a weak magnetic field. That means there is a conversion of
perpendicular to parallel energy. It looks like an acceleration of
particles along the magnetic field caused by a certain force. The
name of this force is magneto-mirror. But this force is a
pseudo-force as there is only a swapping of energy from a
perpendicular into a longitudinal direction, the total energy
doesn't change. If the instability criterion is not fulfilled the
fluid mirror modes show an oscillatory behavior.

Contrary to the fluid description of the mirror instability in the
kinetic description not all particles equally react to the magnetic
field changes \cite{Tajiri}. Particles with small parallel speed
``do not feel'' the mirror force to the same degree as the particles
with larger parallel speed. In this way with changing magnetic field
its energy is not conserved, and the perpendicular pressure changes
synchronously with the magnetic field change. Unlike the fluid
approach in the kinetic treatment the mirror modes become
non-oscillating (exponentially damped) if the instability criterion
is not satisfied. This is because of the resonant origin of the
kinetic mirror instability \cite{South}.

We can compare the growing rate of the mirror instability in our
fluid description Eq. (\ref{gromir}) with the different kinetic
estimates. There is good agreement. Note that the more exact
numerical calculations of the kinetic growing rate for bi-Maxwellian
plasma \cite{Gary93} show that a maximum of the growth rate occurs
at $k_\bot r _ {Bi} \sim 1$, well above the threshold, and $
\Gamma_{\mathrm max} \sim k_\| v_ {Ti \|} $.

We focused our interest on the properties of the mirror modes
because of the high probability of their realization in practice.
The properties of the other modes, fire hose and thermal modes, are
well known. The first one is the prototype of Alfv\'{e}nic waves and
the second one is analogous to the ionic magnetic sound in the
kinetic approach. At places of a crossing of wave branches we have a
mixing of modes, i.e. there is a resonant interaction of wave modes.
As usually in hydrodynamics if we consider the anisotropy of the
plasma as well as its spatial inhomogeneity, these points of mode
intersection will introduce singularities into the wave equations.
Unlike the kinetic approach where wave modes can grow or fade as a
result of a resonance of particles and fluctuations, in the fluid
description it occurs as the result of resonant interaction of
different modes.

In the present paper the linear instability problem is investigated
in a homogenous, unlimited plasma. The classic incompressible fire
hose modes are not influenced by the heat fluxes, and their
instability criterion is the same as that in kinetic theory.
However, the two CGL mirror modes are strongly modified by the heat
fluxes, and there appear two additional thermal branches. These
thermal modes result from including the two dynamic evolution
equations of the thermal fluxes. In the initial state nonzero heat
fluxes are supposed, $\gamma \ne 0$. However, even for zero initial
heat fluxes ($\gamma = 0$) the thermal modes appear. The deduced
8-th order polynomial dispersion equation describes the interaction
of all of the 4 types of compressible modes and their stability. If
the initial heat fluxes along the magnetic field ($\gamma \ne 0$)
are included, the mode dispersion behavior and the instability
criteria are different for the same type of modes running along and
backward with respect to the magnetic field.

To sum it up it can be said that the main shortages of the CGL fluid
theory have been removed. It is shown that the fluid slow mirror
instability criterion is the same as that in the kinetic theory:
$p_\bot^2/p_\| = p_\bot +
 p_{\mathrm m}$. We have shown that in some selected ranges of the
parameters $\alpha$ (parameter of pressure anisotropy),  $\beta$
(parameter of magnetic field), and $\gamma$ (parameter of initial
heat flux) in the plasma there exists such a propagation angle ($l =
\cos^2\theta$) in which at the same time two kinds of fire hose
instability can develop. The discovered new instability is
compressible, slightly periodical (${\mathrm Re}(\omega) \ne 0$),
and it has a larger growing rate than the incompressible fire hose
instability for parallel propagation. It seems that these modes are
analogous to the two kinds of fire hose instabilities found in
kinetic theory \cite{Hollweg70,Pae99,Li00,Hellinger00}. This new
instability develops when the three retrograde modes (two thermal
and fast mirror) interact resonantly.

We found a strong dependence of the growth rate on the parameters
$\alpha, \beta, \gamma$ and $l$. There appear different unstable and
stable wave branches simultaneously within the given parameter
ranges. Only the mode with the highest growth rate will dominate,
and after some exponential growth the nonlinear stage of the
instability should be considered. It is of basic importance for the
found instabilities that in the collisionless plasma there is a
plasma pressure anisotropy that is kinetically supported. The origin
of this pressure imbalance is not important for our fluid
approximation; it is the background of large-scale flows only. In
principle many kinds of kinetic wave turbulence can support such a
pressure anisotropy, the existence of which in the considered plasma
situation is shown by observations.

\subsection{Solar wind}
We think that the discovered wave modes and their instabilities in
the anisotropic fluid approximation are interesting for those
plasmas for which the approximation for a magnetized hot plasma with
rare collisions can be applied. Important candidates for such
conditions are the solar wind and the solar corona plasma.
Macroscopic turbulence observed in the solar wind \cite{Marsch06}
and in the stable coronal turbulent background (appearing in the
nonthermal broadening of coronal emission line profiles
\cite{Asc08}) may be a consequence of these instabilities. Moreover,
it is now generally accepted that the observed large ion temperature
anisotropies are related to the physical mechanism by which the
solar corona and solar wind are heated \cite{Hollweg02,Marsch06}.

Near the Sun heavy ions are stronger heated than protons and
electrons. These findings have strengthened the arguments in favour
of the kinetic ion-cyclotron model of heating and acceleration of
particles in the solar wind. However, this mechanism has a number of
shortages. For example, the observed properties of low-frequency
wave turbulence are close to those of Alfv\'{e}nic modes and their power
spectrum has a maximum around one--two hours. In order to realize
the ion-cyclotron resonance, Hollweg \cite{Hollweg08} assumed that
low-frequency waves by the nonlinear cascade should finally turn to
high-frequency modes. Within the frames of our fluid model such
observed low-frequency modes can easily be explained. Indeed, in the
ideal case (without a heat flux, $\gamma=0$) near the instability
threshold of the Alfv\'{e}nic fire hose modes it is possible to receive
very low frequencies. For the nonideal case (with a heat flux, $
\gamma\ne 0$) the mirror modes have every chance to explain the
observable low frequencies. The new observed facts which can
essentially modify our ideas about the physical nature of the solar
wind are presented in a recent paper \cite{Brovsky}. It appears that
the solar wind consists of sets of magnetic filaments. We think that
the nonlinear evolution of the large-scale mirror modes are capable
of creating such structures.

\subsection{Identification of wave modes}
In the preceding Section 7 we presented the relations between the
fluctuations of the plasma mass density and the magnetic field as
well as the two components of the fluid speed. These formulae can be
used for identifying the modes; in particular the simple analytic
estimates of the asymptotic limiting cases are helpful for
recognizing easily the modes in the observed data. Modern space and
ground-based observations with spectral and imaging methods allow to
detect a reach spectrum of different kinds of wave motions in the
corona \cite{Asc05,Asc08,Marsh09,Wang09}. From such observed data we
get information mainly on the amplitudes of density perturbations
and the fluid velocity component along the line-of-sight, the wave
frequency, the phase speed, the mode life time, and on the spatial
orientation of the magnetic loops along which the waves are running.
These data allow us to compare the observed wave motions with
theoretical predictions \cite{Nakar05}. So far all interpretations
are based on the well developed isotropic MHD wave theory which is
based on the collision-dominated plasma description. However, in
such theoretical interpretations we should be more careful,
especially concerning the outer corona. For example, even in the
lower corona close to the transition region the simulated electron
heat flux is not described correctly by the collision-dominated
Spitzer law \cite{Landi01}. The observed fluxes are closer to the
collisionless theoretical estimates. The appearance of anisotropic
wave modes is an important evidence: if there exists strong enough
heat fluxes, then the phase velocities along the magnetic field will
differ from those in the opposite direction. Observed life times can
be compared with the growing times of the instabilities.

Our theory should be further improved. In the 16-moments transport
equations the next order terms should be retained. Finite particle
gyroradii should result in a $k$-dependence of the maximum of the
growing instability rates such as shown in the kinetic theory
\cite{Lang02}. An inclusion of small collisional terms in the basic
equations would be the best way to describe the coronal plasma.

\begin{acknowledgement}

The present work has been supported by the German Science Foundation
(DFG) under grant No. 436 RUS 113/931/0-1 (R) and by the Russian
Foundation for Basic Research (RFFI) under project No. 09-02-00494
which is gratefully acknowledged.
\end{acknowledgement}


\end{document}